\documentclass{aa}  

\usepackage{graphicx}
\usepackage{txfonts}
\usepackage{natbib}
\usepackage{xspace}
\usepackage{color}
\usepackage{hyperref}

\usepackage{threeparttable}
\usepackage{supertabular}
\usepackage{multirow}
\usepackage{makecell}

\usepackage{amsmath}
\usepackage{subcaption} 
\usepackage{upgreek}
\usepackage{float}

\usepackage{tablefootnote}
\usepackage{tikz}
\usepackage{lscape}
\newcommand{\cref}{\ref}

\begin{document}


\title{The high-energy cyclotron line in 2S~1417-624 discovered with Insight-HXMT during the 2018 outburst}

\author{Q. Liu\inst{1}
\and A. Santangelo\inst{1}
\and L. D. Kong\inst{1}
\and L. Ducci\inst{1}
\and L. Ji\inst{2}
\and W. Wang\inst{3}
\and M. M. Serim\inst{1}
\and C. Güngör \inst{4}
\and Y. L. Tuo\inst{1}
\and D. Serim\inst{1}
}
\institute{Institut für Astronomie und Astrophysik, Universität Tübingen, Sand 1, 72076 Tübingen, Germany \\
\email{qi.liu@mnf.uni-tuebingen.de} \\
\email{andrea.santangelo@uni-tuebingen.de}
\and 
School of Physics and Astronomy, Sun Yat-Sen University, Zhuhai 519082, China
\and
Department of Astronomy, School of Physics and Technology, Wuhan University, Wuhan 430072, China \\ \email{wangwei2017@whu.edu.cn}
\and
İstanbul University, Science Faculty, Department of Astronomy and Space Sciences, Beyazt, İstanbul 34119 Turkey 
}
\date{ }

\abstract
{We report a detailed timing and spectral analysis of the X-ray pulsar 2S~1417-624 using the data from Insight-HXMT during the 2018 outburst. The pulse profiles are highly variable with respect to both unabsorbed flux and energy. A double-peaked pulse profile from the low flux evolved to a multi-peaked shape in the high-flux state. The pulse fraction is negatively correlated to the source flux in the range of $\sim$(1--6)$\ \times \ 10^{-9}$ erg cm$^{-2}$ s$^{-1}$, consistent with \textit{Rossi} X-ray Timing Explorer (RXTE) studies during the 2009 giant outburst. The energy-resolved pulse profiles around the peak outburst showed a four-peak shape in the low-energy bands and gradually evolved to triple peaks at higher energies. The continuum spectrum is well described by typical phenomenological models, such as the cut-off power law and the power law with high-energy cut-off models. Notably, we discovered high-energy cyclotron resonant scattering features (CRSFs) for the first time, which are around 100 keV with a statistical significance of $\sim$7$\sigma$ near the peak luminosity of the outburst. This CRSF line is significantly detected with different continuum models and provides very robust evidence for its presence. Furthermore, pulse-phase-resolved spectroscopy confirmed the presence of the line, whose energy varied from 97 to 107 keV over the pulse phase and appeared to have a maximum value at the narrow peak phase of the profiles.
}


\maketitle

\section{Introduction}
\label{sec:introduction}

Accreting X-ray binary pulsars (XRBPs) were first discovered with the Uhuru satellite. They are highly magnetized neutron stars in binary systems that accrete matter from an optical companion. Most of these sources belong to the class of Be/X-ray binaries, where the optical star is a star of spectral type B with emission lines \citep{1976Natur.259..292M}. Be/X-ray binaries are known to exhibit Type I and Type II outbursts \citep{2011Ap&SS.332....1R}. Type I outbursts usually occur periodically around the periastron passage of the neutron star and last for 0.2--0.3 of the orbital period with a peak luminosity of $\leq 10^{37}$ erg s$^{-1}$. Type II outbursts occur occasionally and do not depend on orbital phases. They last for a large segment of the orbital period or even for several orbital cycles, and their peak luminosity reaches up to $\sim$$10^{38}$ erg s$^{-1}$ \citep{2011Ap&SS.332....1R}. 

The matter that is accreted by the neutron stars is channelled onto the magnetic poles by the magnetic field. The X-ray emissions are produced in the hot spots or accretion columns at the magnetic poles, which originate from the released kinetic energy of falling materials. There are two typical accretion regimes, depending on the mass-accretion rates or accretion luminosity. At low mass-accretion rates, that is, in the sub-critical regime, the radiation pressure can be neglected, and no radiation-dominated shock is expected to form. The falling plasma is decelerated either via Coulomb interactions, resulting in hot spots on the surface of the neutron star, or via a collisionless shock above the neutron star surface (see e.g. \citealt{2022arXiv220414185M}). At high mass-accretion rates in contrast, that is, in the super-critical regime, a radiation-dominated shock appears above the neutron star surface \citep{2012A&A...544A.123B,2015MNRAS.447.1847M}. The transition between the two regimes is crucial for studying the complex interaction between the magnetic field and the accretion flow. The geometry of the emission region also depends on the mass-accretion rate, which is reflected in the variations in the pulse profile shapes.

The source 2S~1417-624, discovered with the Third Small Astronomy Satellite (SAS-3) in 1978 \citep{Apparao1980}, is a transient Be/X-ray binary that showed giant Type II outbursts and a number of small Type I outbursts. It has a pulsation of $\sim$17.64 s \citep{1981ApJ...243..251K} with a spin-up rate of $\sim$(3--6) $\times$ $10^{-11}$ Hz s$^{-1}$ \citep{Finger1996}, and it even shows a spin-down trend during quiescence, as observed by \cite{2010MNRAS.406.2663R}. The orbital period and eccentricity are measured to be about 42.12 d and 0.446  \citep{Finger1996}. The optical companion has been recognized as a B1 Ve star, with a distance of 1.4--11.1 kpc estimated by \cite{1984ApJ...276..621G}. Recently, the distance was estimated to be $\sim$9.9 kpc based on Gaia data \citep{Bailer2018}. After its discovery, several giant outbursts of the source were observed and studied, such as the outburst in 1994, which was detected by the Burst and Transient Source Experiment (BATSE) \citep{Finger1996}, and the outburst in 1999, which was detected by the RXTE \citep{Inam2004}. During a giant outburst in 2009, \cite{2018MNRAS.479.5612G} reported based on data from RXTE that the pulse profiles evolved from a double-peaked shape at lower luminosity to a triple-peaked shape around peak luminosity, and the pulse fraction was anti-correlated with the source flux. The authors concluded that the beam configurations are associated with the transition from sub-critical to super-critical regimes. This anti-correlation between the pulse fraction and luminosity was also reported for the 2018 outburst, which was observed by NuSTAR \citep{2019MNRAS.490.2458G}. The source 2S~1417-624 has also been observed by \textit{Chandra} in the quiescent state in 2013 May \citep{2017MNRAS.470..126T}, and it was reported that the source spectrum can be modelled by either a power law or a blackbody function around a temperature of 1.5 keV. The detected giant outburst in 2018 reached a peak intensity of $\sim$350 mCrab as recorded in \textit{Swift}/BAT in the 15–50 keV energy band. During the 2018 outburst, pulse profiles evolving into quadruple-peaked features with increasing luminosity were observed by NICER \citep{2020MNRAS.491.1851J}. \cite{2022MNRAS.510.1438S} confirmed the transition from a sub-critical to a super-critical regime in terms of spectral changes in the source in the 2018 outburst with NICER. The detailed evolution of the pulse profiles or pulse fractions in a broader energy band for the luminosity in this giant outburst has not been reported so far, however, but it is also essential to report this. 

It is known that the CRSFs can provide a direct estimate of the magnetic strength of the pulsar. However, no cyclotron line has been detected in the pulsar spectra of 2S~1417-624 in the 0.9–79 keV range so far \citep{2019MNRAS.490.2458G}, not even at the peak of the giant outburst. \citep{2020MNRAS.491.1851J} estimated the magnetic field of 2S~1417-624 to be $\sim$$7 \times 10^{12}$G and reported a distance of $\sim$20~kpc after considering the accretion torque effect. Based on the wide energy band and large effective area at high energy, we study the observations of the giant 2018 outburst performed by Insight-HXMT (\citealt{2020SCPMA..6349502Z}), which currently is the brightest outburst, even though the source recently underwent an outburst in 2021, which was monitored by Fermi/GBM and \textit{Swift}/BAT. In Sect. \cref{sec:Data} we briefly describe the Insight-HXMT observations and data reduction. We report the timing and spectral studies, including the phase-averaged and phase-resolved analyses, in Sect. \cref{sec:results}. In Sect. \cref{sec:discussion} we estimate the magnetic field by CRSFs and discuss the theoretically expected critical luminosity.

\section{Observations and data reduction} \label{sec:Data}

\begin{figure}
    \centering
    \includegraphics[width=.5\textwidth]{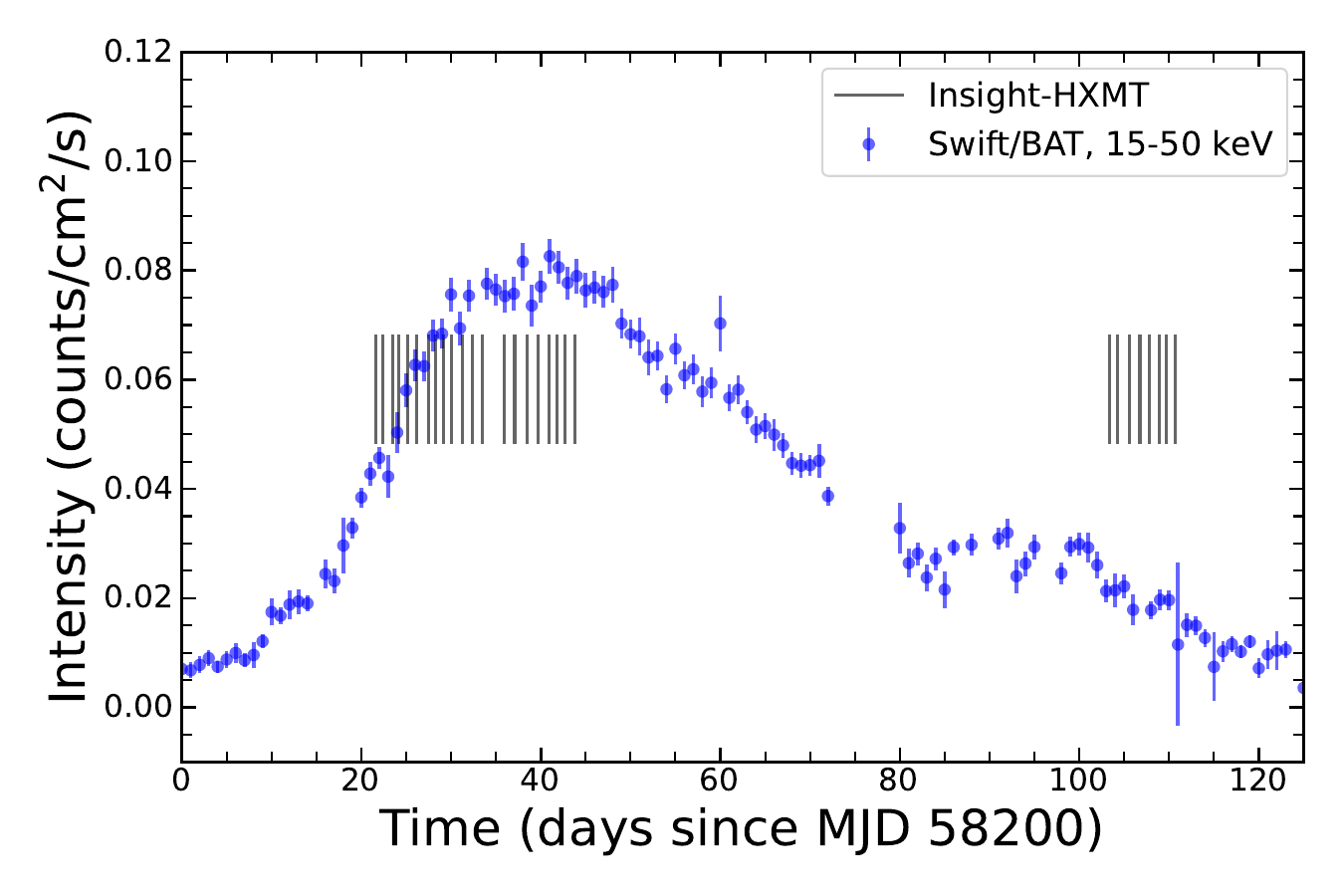}
    \caption{Light curve of the 2018 outburst via \textit{Swift}/BAT in the energy range of 15-50 keV (blue dots) with the observation times of the Insight-HXMT data (vertical black lines).}
    \label{fig:counts}
\end{figure}

The observations of 2S~1417-624 presented in this paper were made with Insight-HXMT from April 13, 2018, until July 11, 2018. A total of 29 pointed observations during the 2018 outburst were used. They are marked as vertical black lines in Fig. \cref{fig:counts}. Table \cref{tab:ObsIDs} provides details of the observational information. The Insight-HXMT satellite is equipped with three scientific payloads: the High Energy X-ray telescope (HE; 20--250 keV), the Medium Energy X-ray telescope (ME; 5--30 keV), and the Low Energy X-ray telescope (LE; 1--15 keV). Their effective areas are 5000 $\rm cm^2$, 952 $\rm cm^2$, and 384 $\rm cm^2$, respectively, with time resolutions of 25 $\mathrm{\mu}$s, 276 $\rm \mu s$, and 1 ms for HE, ME, and LE, respectively.

For the data reduction, the Insight-HXMT Data Analysis Software (HXMTDAS v2.04) was employed. We calibrated, screened, and extracted high-level products from the event files using standard criteria suggested by the HXMTDAS user guide. The tasks \textsl{he/me/legtigen} were used to generate a good-time interval (GTI) file with pointing offset angles $<0.04^\circ$, elevation angles $>10^\circ$, geomagnetic cut-off rigidity $>$ 8 GeV, and eliminated intervals within the South Atlantic Anomaly passage. The tasks \textsl{he/me/lescreen} were employed to screen the data. The arrival times of events were then corrected to the Solar System barycenter using the task \textit{hxbary}. The tasks \textit{he, me, and lelcgen} were employed to extract light curves with a bin size of 0.0078125 (1/128) sec. For the spectral analysis, the tasks \textsl{he, me, and lespecgen} were used to generate the spectra, with 2–10 keV for LE, 10–30 keV for ME, and 30–140 keV for HE. The tools \textsl{he, me, and lebkgmap} can be employed to estimate the background. XSPEC 12.11.1 \citep{1996ASPC..101...17A} was used for the spectrum-fitting analysis. All uncertainties were estimated using a Markov chain Monte Carlo (MCMC) method with a chain length of 20,000 steps, which are at the 68\% confidence level unless otherwise noted.

\begin{figure*}
    \centering
    \includegraphics[width=.99\textwidth]{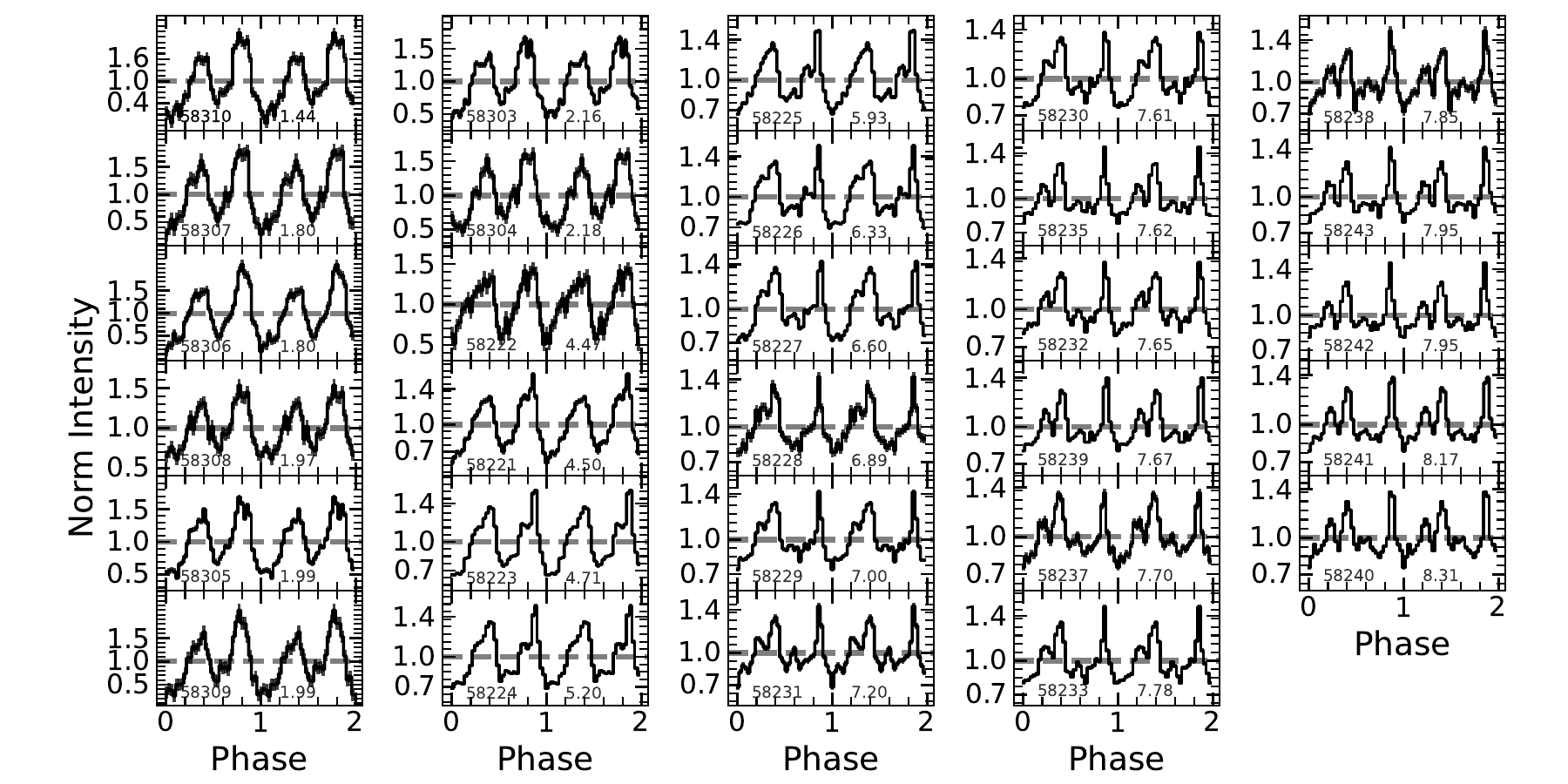}
    \caption{Pulse proﬁles in the energy range of 30–100 keV from Insight-HXMT/HE data at different times (in MJD, left in each panel) and flux (in $10^{-9}$ erg cm$^{-2}$ s$^{-1}$, right in each panel) during the 2018 giant outburst.}
    \label{fig:pulse_HE}
\end{figure*}

\begin{figure*}
    \centering
    \includegraphics[width=.49\textwidth]{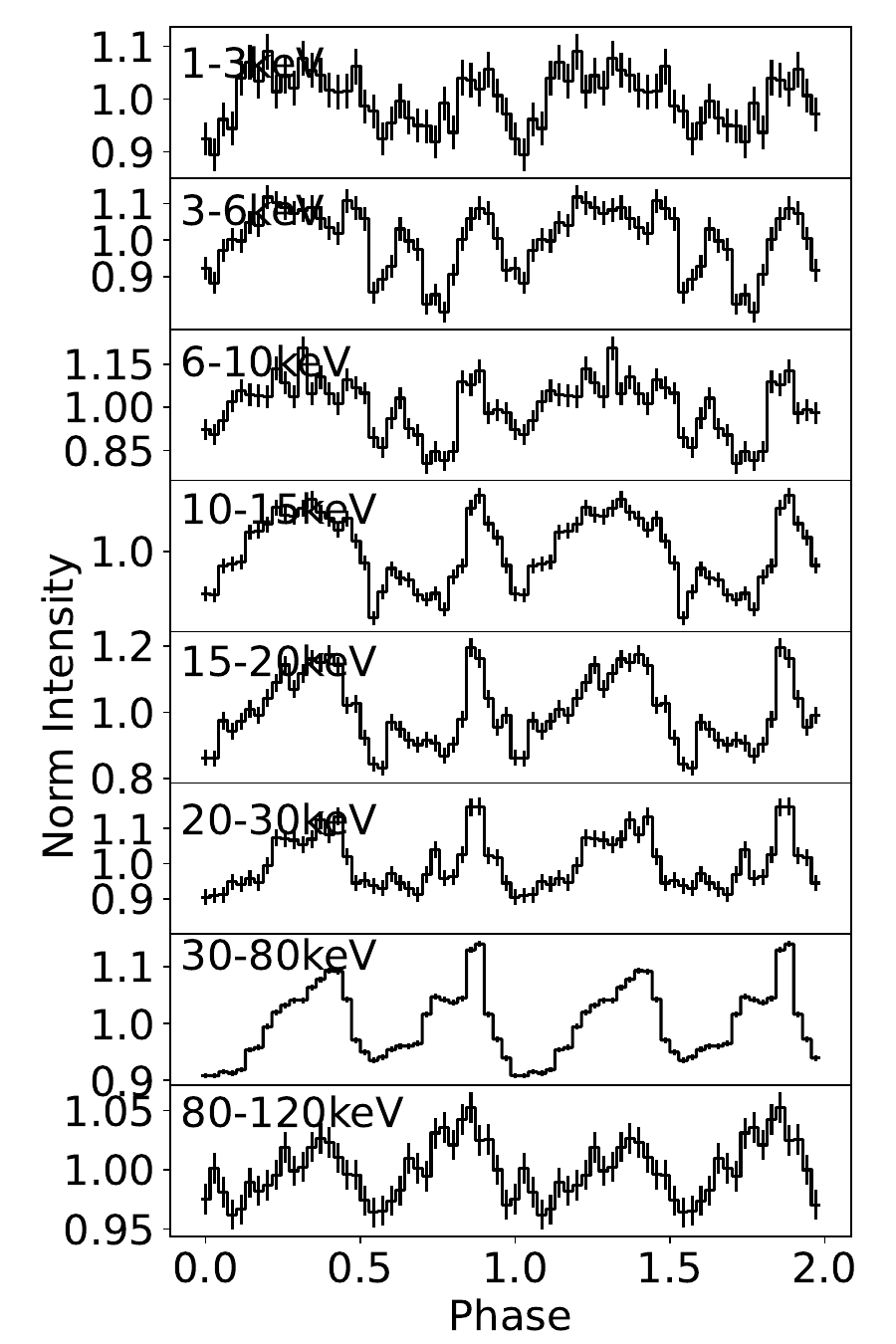}
    \includegraphics[width=.49\textwidth]{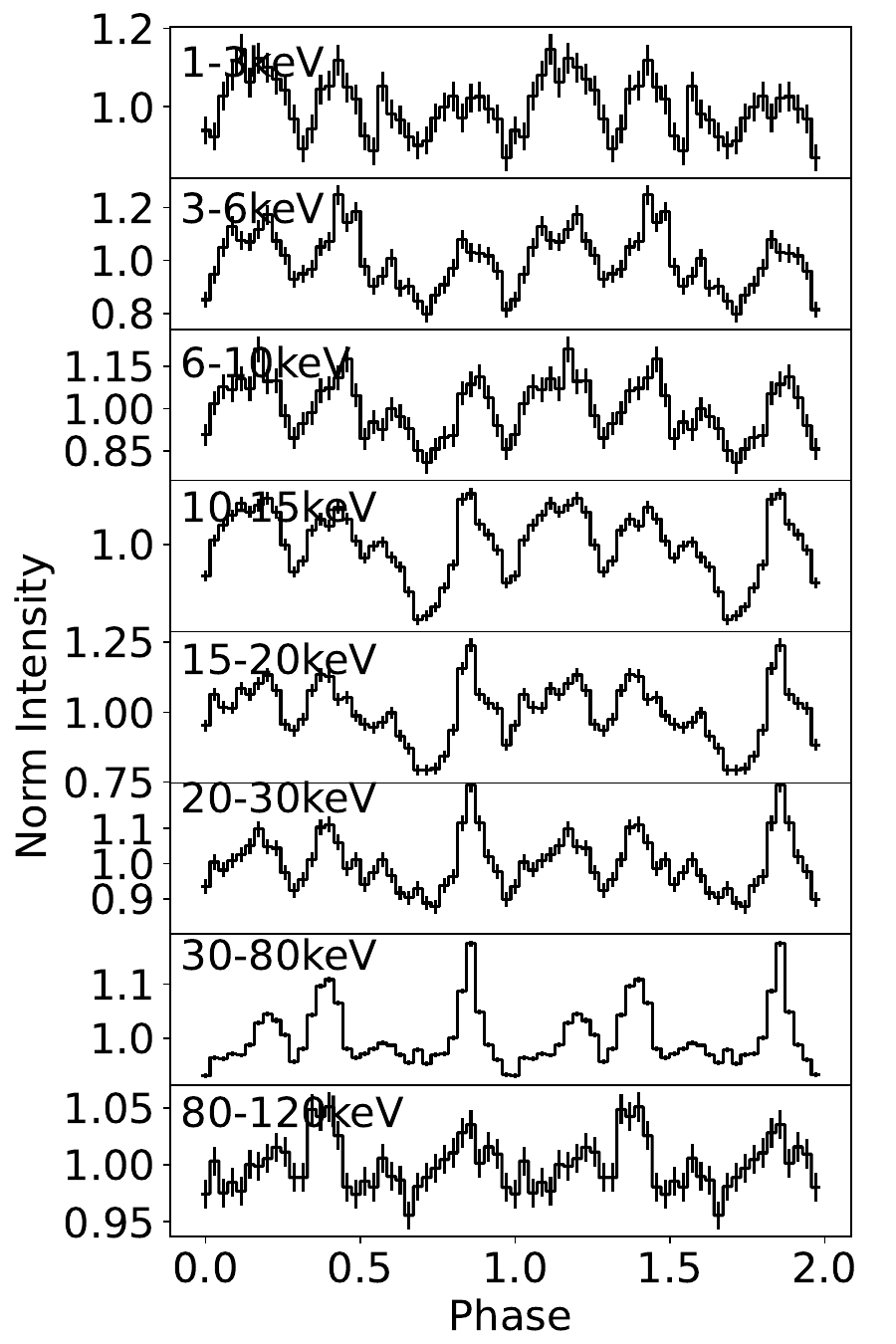}
    \caption{Energy-resolved pulse proﬁles of 2S~1417-624 for ObsID P0114709003 in the rising stage of the 2018 outburst (left panels, low luminosity) and ObsID P0114709020 near the peak of the outburst (right panels, high luminosity) based on Insight-HXMT.}
    \label{fig:profile_energy}
\end{figure*}

\begin{figure}
    \centering
    \includegraphics[width=.5\textwidth]{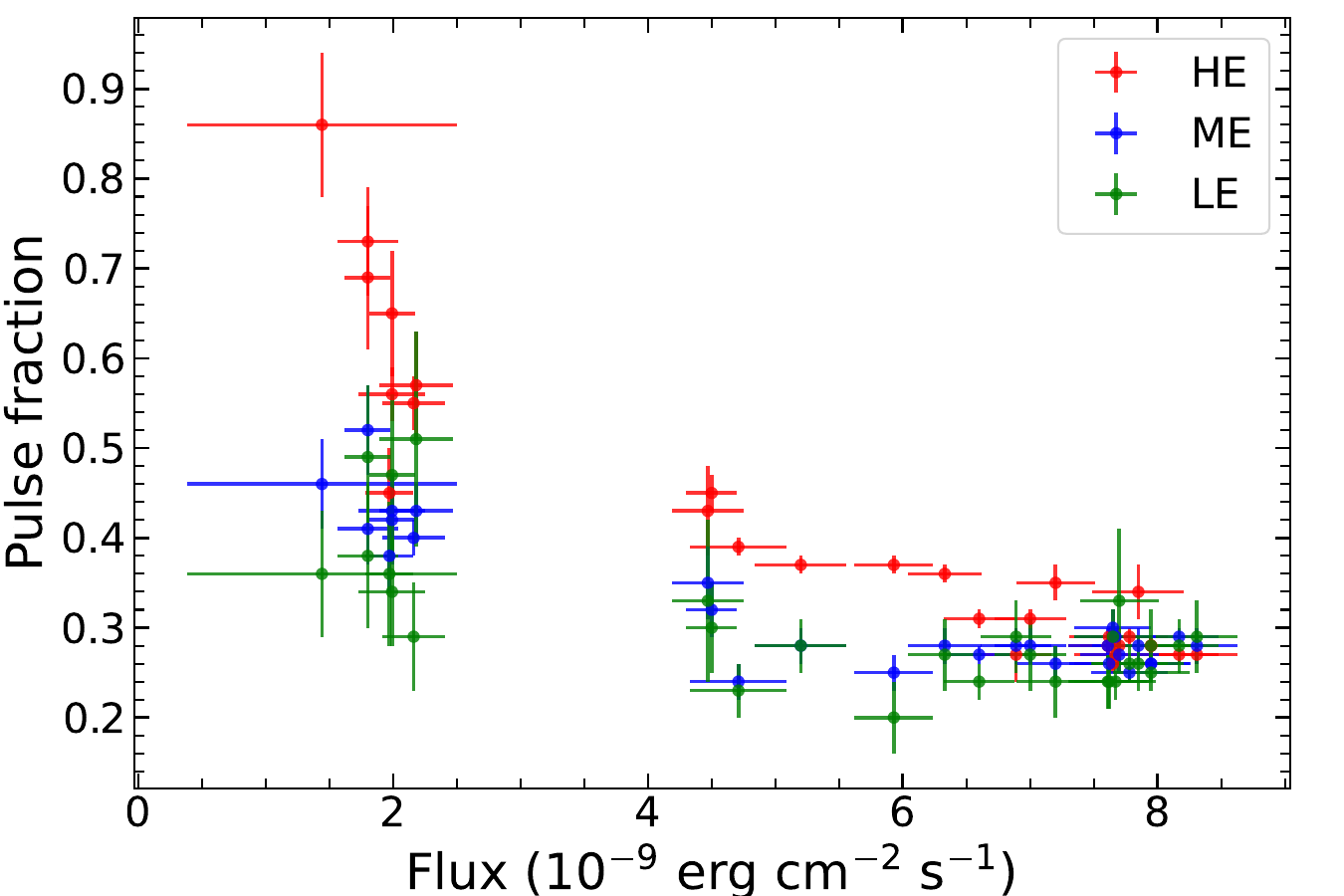}
    \caption{Pulse fraction as a function of flux in the HE (30--60 keV), ME (10--30 keV), and LE (1--10 keV) energy bands (also see Table \cref{tab:ObsIDs}).}
    \label{fig:pf_energy}
\end{figure}

\section{Results} \label{sec:results}

\begin{table*}
\centering
\caption{Best-fitting parameters in different models.}

\large
\renewcommand\arraystretch{1.23}
\setlength{\tabcolsep}{6.5mm}{
\begin{tabular}{l|rr|rr}
\hline
& \multicolumn{2}{c}{cut-off PL} & \multicolumn{2}{c}{HIGHECUT} \\
\cline{2-5}
& gabs$^1$ & cyclabs$^2$ &  gabs & cyclabs\\

\hline
 $E_{\rm cyc}$ (keV) & $102.7_{-3.6}^{+4.7}$ & $99.5_{-3.3}^{+4.2}$  &  $102.5_{-3.7}^{+4.5}$ &   $99.7_{-2.0}^{+2.8}$  \\
 $\sigma_{\rm cyc}$ (keV)/$W_{\rm f}$ (keV) &  $11.7_{-2.8}^{+3.7}$ &  $12.6_{-4.7}^{+4.6}$ &  $12.0_{-2.4}^{+3.3}$  &  $11.6_{-3.5}^{+3.0}$ \\
 $d_{\rm cyc}$ (keV)/$D_{\rm f}$ & $14.2_{-3.6}^{+3.5}$ &  $0.6_{-0.1}^{+0.2}$  &  $14.7_{-3.3}^{+4.3}$ &  $0.6_{-0.1}^{+0.2}$ \\
 $N_{H}$  &$1.6_{-0.2}^{+0.2}$ &  $1.7_{-0.3}^{+0.3}$ & $2.4_{-0.3}^{+0.3}$ &  $2.5_{-0.3}^{+0.3}$ \\
 $\Gamma$ & $0.21_{-0.01}^{+0.01}$ &  $0.22_{-0.01}^{+0.01}$  &  $0.21_{-0.01}^{+0.01}$ & $0.22_{-0.01}^{+0.01}$ \\
 norm & $0.033_{-0.001}^{+0.001}$ &  $0.034_{-0.001}^{+0.001}$ &  $0.027_{-0.001}^{+0.001}$ & $0.027_{-0.001}^{+0.001}$ \\
$E_{\rm cut}$(keV) &  $16.8_{-0.2}^{+0.2}$ &  $17.1_{-0.2}^{+0.1}$ &  $16.9_{-0.1}^{+0.2}$  &   $17.1_{-0.2}^{+0.1}$ \\
$E_{\rm f}$(keV) &  ... & ... &  $4.1_{-0.3}^{+0.2}$   &  $4.1_{-0.2}^{+0.2}$  \\
$E_{\rm Fe}$(keV) & $6.48_{-0.03}^{+0.03}$  & $6.47_{-0.02}^{+0.03}$  &  $6.47_{-0.03}^{+0.03}$  &  $6.46_{-0.03}^{+0.04}$ \\
$\sigma_{\rm Fe}$(keV) & $0.16_{-0.03}^{+0.03}$  & $0.15_{-0.02}^{+0.03}$   &   $0.15_{-0.05}^{+0.04}$   &   $0.13_{-0.05}^{+0.05}$  \\
$I_{\rm Fe}$(keV) & $0.0009_{-0.0001}^{+0.0001}$ & $0.0009_{-0.0001}^{+0.0001}$  &  $0.0009_{-0.0002}^{+0.0002}$   & $0.0009_{-0.0001}^{+0.0001}$ \\
$kT_{\rm bb}$ (keV) & $0.57_{-0.02}^{+0.02}$ &  $0.56_{-0.02}^{+0.01}$ & $0.53_{-0.01}^{+0.01}$   &  $0.53_{-0.01}^{+0.02}$ \\
Cons$_{\rm ME/LE}$ & $1.03_{-0.01}^{+0.01}$ & $1.03_{-0.01}^{+0.01}$ &  $1.03_{-0.01}^{+0.01}$  &  $1.03_{-0.01}^{+0.01}$  \\
Cons$_{\rm HE/LE}$ & $1.01_{-0.01}^{+0.01}$  &  $1.00_{-0.01}^{+0.01}$ &  $1.00_{-0.01}^{+0.01}$   &  $1.01_{-0.01}^{+0.01}$ \\
reduced $\chi^2$ (dof) & 1.04 (1345) & 1.04 (1345) & 1.04 (1344)   &  1.04 (1344) \\ \hline
$\rm Flux_{2-140}$ & \multicolumn{4}{c}{$7.89_{-0.02}^{+0.02}$}  \\
\hline
\end{tabular}
}
\tablefoot{
The flux is given in units of $10^{-9}$ erg cm$^{-2}$ s$^{-1}$, and the column density $N_{\rm H}$ is listed in units of 10$^{22}$ atoms cm$^{-2}$. For the normalization constants, we fixed the constant to one for the LE instrument and kept the Cons$_{\rm ME/LE}$ and Cons$_{\rm HE/LE}$ free. \\
$^1$ gabs(E)$=\exp \left(-\frac{d_{\mathrm{cyc}}}{\sqrt{2 \pi} \sigma_{\mathrm{cyc}}} e^{-0.5 \left[\left(E-E_{\mathrm{cyc}}\right) / \sigma_{\mathrm{cyc}}\right]^2}\right)$ \\
$^2$ cyclabs(E)$=\exp \left[-D_{\mathrm{f}} \frac{\left(W_{\mathrm{f}} E / E_{\mathrm{cyc}}\right)^{2}}{\left(E-E_{\mathrm{cyc}}\right)^{2}+W_{\mathrm{f}}^{2}}\right]$
}
\label{tab:pars}
\end{table*}

\subsection{Timing studies}

To determine the pulse period, we used the epoch-folding technique with the task \textit{efsearch} in ftools\footnote{A General Package of Software to Manipulate FITS Files}. We searched for pulsations by folding the light curve in the energy range of 10--30~keV with 50 phase bins per period around the initial spin period of 17.48 s, consistent with the history of the pulse frequency measured using Fermi/GBM\footnote{\url{https://gammaray.nsstc.nasa.gov/gbm/science/pulsars/lightcurves/2s1417.html}}. The best pulse period was identified based on the maximum $\chi^{2}$ value. The pulse profiles in 30--100~keV, 10--30~keV, and 1--10~keV energy bands were generated by folding each light curve using the best spin period. These pulse profiles were arranged as the flux rises and exhibit significant changes in different flux states. As depicted in Fig. \cref{fig:pulse_HE}, the pulse profile in 30--100 keV for Insight-HXMT/HE appeared to be broad double peaks at a low flux level (also see \cite{Finger1996,Inam2004}), which is consistent with the Insight-HXMT observation of the 25--100~keV at the low intensity reported by \cite{2020MNRAS.491.1851J}. As the intensity towards the peak flux increased (>4.5$\ \times \ 10^{-9}$ erg cm$^{-2}$ s$^{-1}$), the peak in the 0.8 phase became narrower. Additionally, another peak gradually split into two peaks, and a triple-peaked profile appears at a flux level of $\sim$7$\ \times \ 10^{-9}$ erg cm$^{-2}$ s$^{-1}$ in the 2–140 keV range. As shown in Fig. \cref{fig:pulse_ME} for the 10–30 keV profiles observed with Insight-HXMT/ME, they behave in a similar way as the pulse profile at 30--100 keV and evolve from a double-peaked profile at low intensity into a four-peak profile in the high state (a flux level of >7$\ \times \ 10^{-9}$ erg cm$^{-2}$ s$^{-1}$). The additional peak in phase 0.6 is more likely caused by the narrower peaks in phase 0.8 and the deeper minimum (dip) in phase 0.5, and it therefore produces a four-peak pulse profile. The 30--100 keV profiles also have the same trend. Figure \cref{fig:pulse_LE} shows pulse profiles at 1--10 keV from Insight-HXMT/LE data. Although the statistics in LE data is low because the exposure time is much shorter and the effective area is smaller than for the ME and HE telescopes of Insight-HXMT, we still note that they show double peaks at a low flux of <4.5$ \ \times \ 10^{-9}$ erg cm$^{-2}$ s$^{-1}$ and triple peaks at an intermediate flux of around (4.5--6.5)$\ \times \ 10^{-9}$ erg cm$^{-2}$ s$^{-1}$. This result is consistent with the transitional flux of $\sim$3.2$\ \times \ 10^{-9}$ erg cm$^{-2}$ s$^{-1}$ reported by \cite{2018MNRAS.479.5612G} in the 3–30 keV range with the RXTE observation in 2009. The authors observed that a double-peaked profile evolved into a triple-peaked profile. Additionally, the pulse profiles of 1--10 keV display four peaks in a higher flux state. Based on the changes in the pulse profiles observed in different energy bands with the flux, it is clear that the evolution of the pulse profiles from double to multiple peaks at the critical flux level of $\sim$7$\ \times \ 10^{-9}$ erg cm$^{-2}$ s$^{-1}$ is associated with the transitions from the sub-critical to super-critical regime. Other Be/X-ray binaries, such as EXO 2030+375 \citep{2016A&A...593A.105F,2017MNRAS.472.3455E}, also showed multiple peaks during outbursts. 

\begin{figure*}
    \centering
   \includegraphics[width=.49\textwidth]{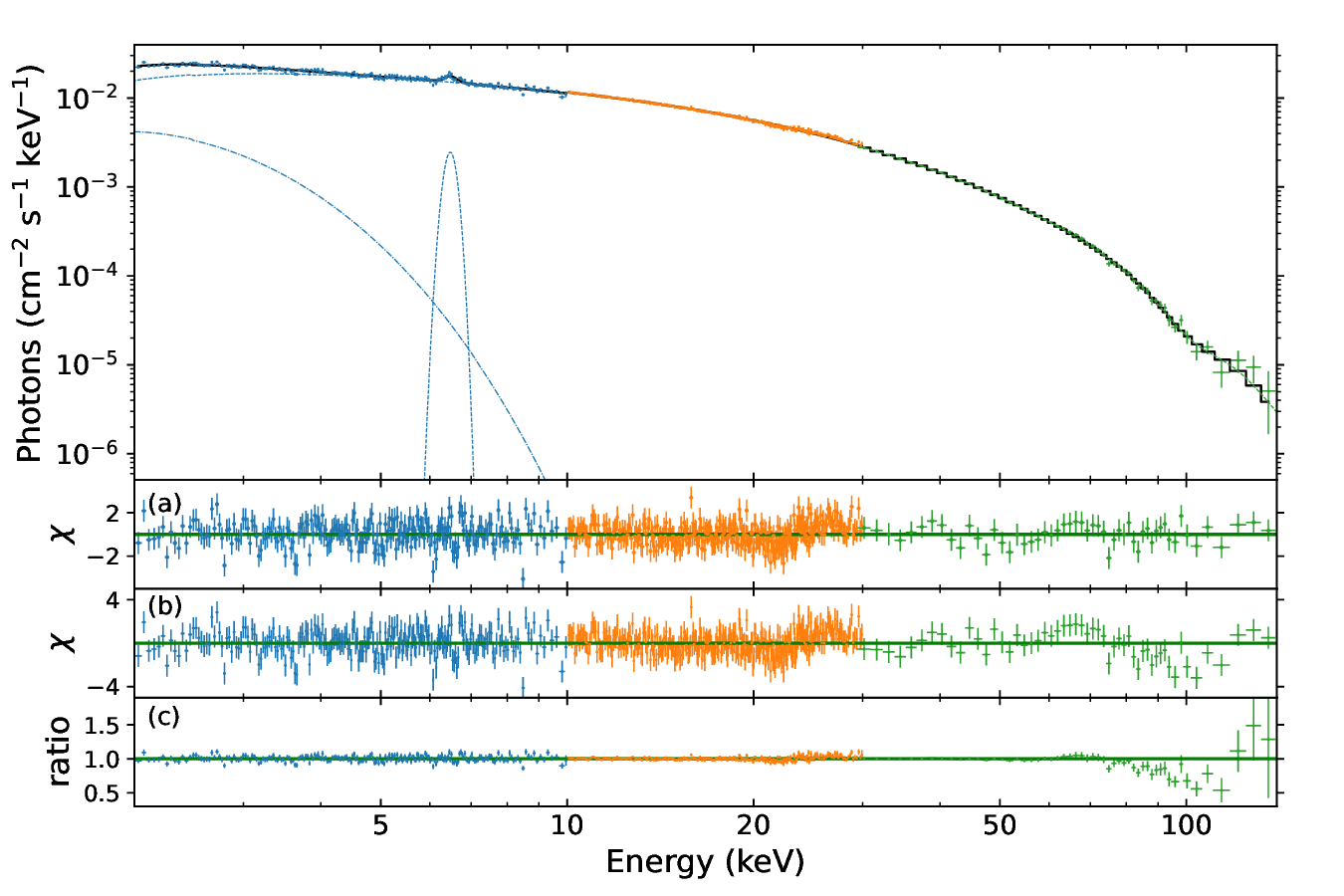}
   \includegraphics[width=.49\textwidth]{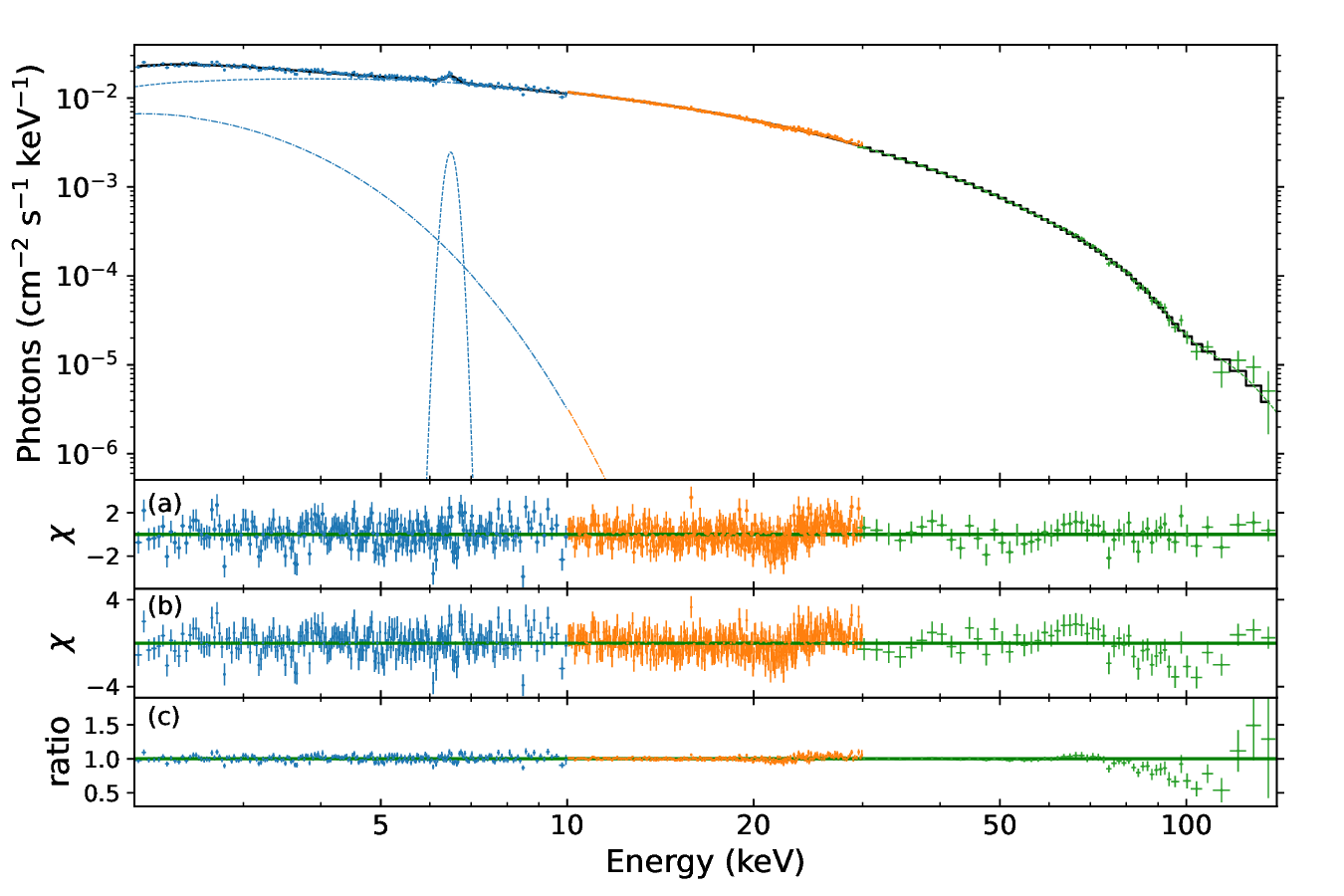}
    \caption{Spectrum of 2S~1417-624 from 2–140 keV fitted by the cut-off PL with the gabs model (left plots) and the HIGHECUT with the gabs model (right plots). Panel a: Residuals $\chi$ ((data-model)/error) of the best-fitting model. Panel b: Residuals for the best-fit model without the 100 keV CRSF. Panel c: Residuals in terms of the data-to-model, ratio setting the strength of the 100 keV cyclotron line to zero.}
    \label{fig:spectrum}
\end{figure*}

To investigate the dependence of the pulse profile on energy during the 2018 outburst, we present the pulse profile for the observation of P0114709003 (MJD 58223.54) during the rising stage of the outburst (low luminosity) and the observation of P0114709020 (MJD 58242.66) near the peak of the outburst (high luminosity) as examples. As shown in Fig. \cref{fig:profile_energy}, the pulsed profile was detected in energies up to 120 keV and showed a significant evolution over different energy bands and luminosities. At low luminosity (left panel), a triple-peaked profile is clearly visible below 20 keV, which then evolves into double peaks at higher energies, consistent with the energy evolution of the pulse profile observed in the 2009 RXTE observation near the peak outburst \citep{2018MNRAS.479.5612G}. However, at high luminosity (right panel), a quadruple-peaked profile was observed in the low energy range of 1--20 keV, while at higher energies, the pulse had a triple-peaked shape, suggesting a different emission geometry compared to the profiles at low luminosity. Additionally, the peak in the 0.8–0.9 phase becomes narrower with increasing energy.

\begin{figure}
    \centering
   \includegraphics[width=.49\textwidth]{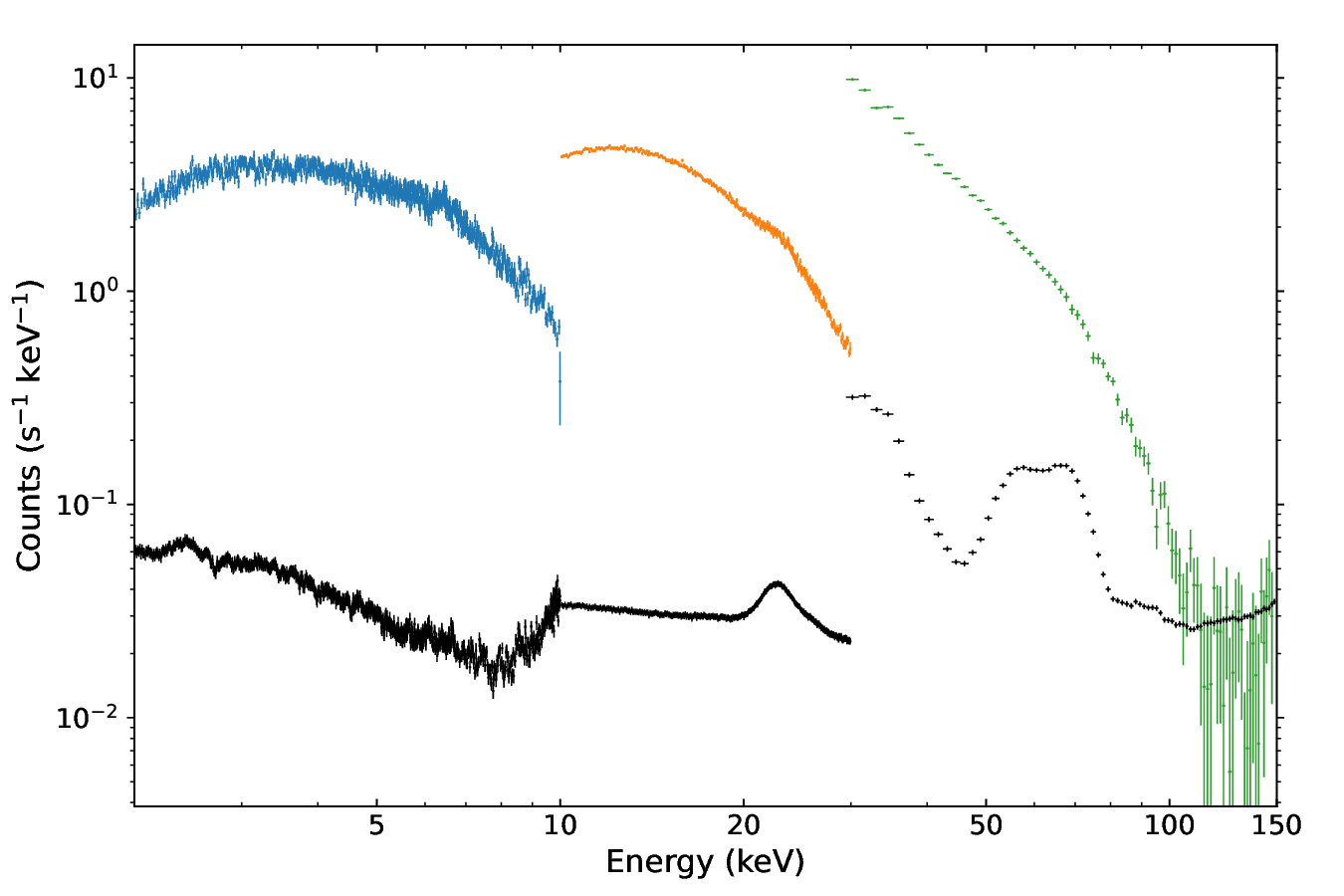}
    \caption{Folded phase-averaged spectra from the LE, ME, and HE. The black line shows the background spectrum with systematic errors of 3.5\%, 2.0\%, and 2.5\% for LE, ME, and HE, respectively, based on the Insight-HXMT background models (\citealt{2020JHEAp..27...24L,2020JHEAp..27...44G,2020JHEAp..27...14L}). The statistical errors of the background are ignored.}
    \label{fig:back}
\end{figure}

We also present the pulse fraction versus the corresponding flux. We found that the pulse fraction in different energy bands decreases as the flux rises. This anti-correlation between pulse fraction and flux was also reported in previous research \cite{2018MNRAS.479.5612G,2019MNRAS.490.2458G}. The red points in Fig. \cref{fig:pf_energy} show the pulse fraction of 30--60 keV and have a clear decreasing trend with source flux. The pulse fraction in the HE bands is larger than in ME or LE, indicating a highly pulsed component with increasing energy, similar to most pulsars (see \citealt{2009AstL...35..433L}). The negative correlation between the pulse fraction and the flux in the 10–30 keV and 1--10 keV ranges is also clearly visible, but above the transitional flux of $\sim$(6--7)$\times 10^{-9}$ erg cm$^{-2}$ s$^{-1}$, it is flat and has a positive trend. This phenomenon can be explained by taking the emission geometry into account. It is well known that the geometry of the emitting region is related to the mass-accretion rate and magnetic field strength. Changes in pulse profiles with varying mass-accretion rates may imply different geometries of the emitting regions, and therefore, different beam patterns. Below the critical flux (sub-critical), the emission geometry is expected to form a pencil beam on the hot spots at the neutron star surface (see \citealt{2022arXiv220414185M}), exhibiting double peaks and highly pulsed radiation. As the accretion rate approaches the critical flux, a radiation-dominated accretion column structure forms \citep{1976MNRAS.175..395B,2007ApJ...654..435B,2015MNRAS.447.1847M}, leading to a transition from a pencil beam to a mixture of pencil and fan beams \citep{2012A&A...544A.123B}. This results in multiple peaks and low pulsed radiation, and therefore, in the observed anti-correlation between pulse fraction and flux. Above the critical flux (super-critical), a full fan-beamed emission pattern forms, and the pulse fraction is expected to be stable or enhanced. The transition from the sub-critical to the super-critical accretion regime may indicate the change from a pencil beam to a fan beam pattern, which is consistent with the evolution of pulse profiles on flux, as seen in our results. However, the fan beam caused by the accretion column structure is a rather simplified picture and reflects our limited understanding of the emission geometry. In reality, the models of accretion column physics are highly complex (see \citealt{2007ApJ...654..435B}, \citealt{2015MNRAS.454.2539M} and \citealt{2021A&A...656A.105T}). Additionally, relativistically downward-beamed and reflected radiation can also contribute to the pulse morphology \citep{2015MNRAS.452.1601P,2013ApJ...777..115P}. Furthermore, the possibility of a complex, non-dipolar magnetic field configuration, leading to the formation of multiple accretion columns, is another scenario that could explain the evolution of pulse profiles.

\subsection{Spectral properties}
\subsubsection{Phase-averaged spectroscopy} \label{sec:averaged}

Since the 2018 outburst was the brightest giant outburst of the source so far, we were able to perform a spectral analysis with these data. To improve the statistics and S/N of the data, we obtained a merged spectrum by combining the nine adjacent observations from ObsID P0114709012 to P0114709021 (the observation of P0114709015 was excluded because the statistics was poor) around the peak luminosity (>7.5 $\times \ 10^{-9}$ erg cm$^{-2}$ s$^{-1}$) using the tasks \textit{addspec} and \textit{addrmf} contained in the HEASARC’s packages. We noted that the pulse profiles for these combined observations are almost identical and the spectral shapes do not change significantly, and therefore, the
superposition of different spectra will not lead to artificial features. The spectrum of 2S 1417-6124 has been found to be well fitted by a cut-off power law (cut-off PL in XSPEC) \citep{2018MNRAS.479.5612G,2019MNRAS.490.2458G,2020MNRAS.491.1851J} or a high-energy cut-off power law (HIGHECUT in XSPEC) continuum model \citep{2022MNRAS.510.1438S,2022Ap&SS.367..112M}, along with a Gaussian component for the 6.4 keV iron emission line. We therefore initially fitted the continuum of the source with the two typical models, the cut-off PL and HIGHECUT. We found that both models could describe the continuum spectra well for the accreting pulsar 2S~1417-624. Additionally, a thermal blackbody component with a temperature of $\sim$1 keV was taken into account, following previous studies \citep{2019MNRAS.490.2458G,2022Ap&SS.367..112M}. For the interstellar absorption, we included the tbabs model in XSPEC and adopted the {\tt wilm} abundances \citep{2000ApJ...542..914W}. 

\begin{figure}
    \centering
    \includegraphics[width=.49\textwidth]{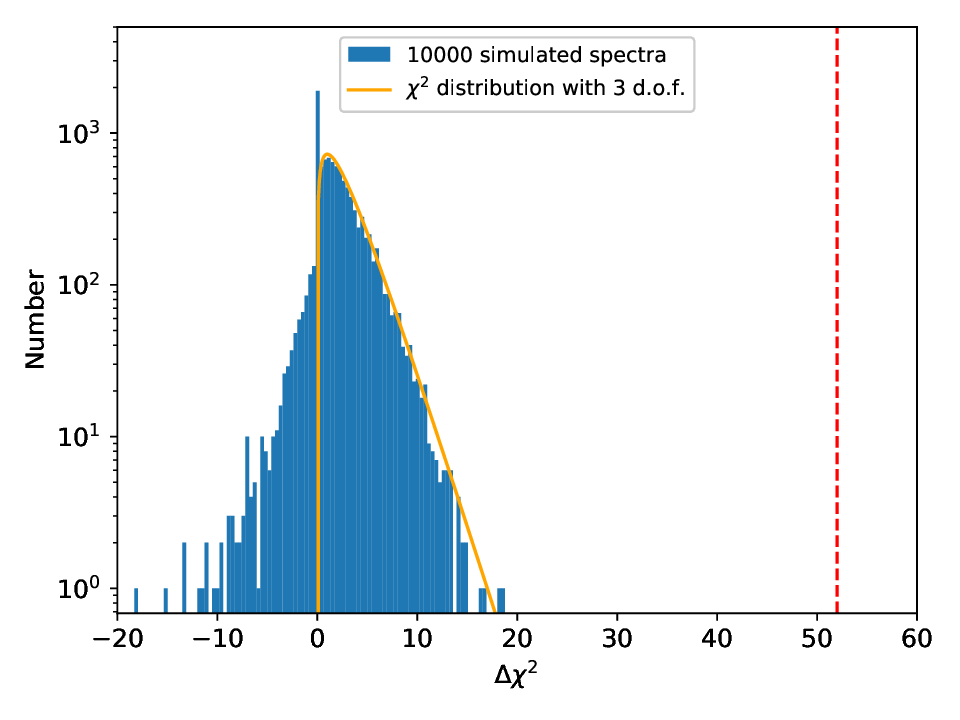}
    \caption{Hstogram from 10000 simulated spectra showing $\Delta \chi^2$ values for the significance simulations. The orange curve shows the $\chi^2$ distribution with 3 d.o.f. The dashed red line shows the observed change in $\Delta \chi^2$ in real data.}
    \label{fig:simul}
\end{figure}

\begin{figure}
    \centering
    \includegraphics[width=.49\textwidth]{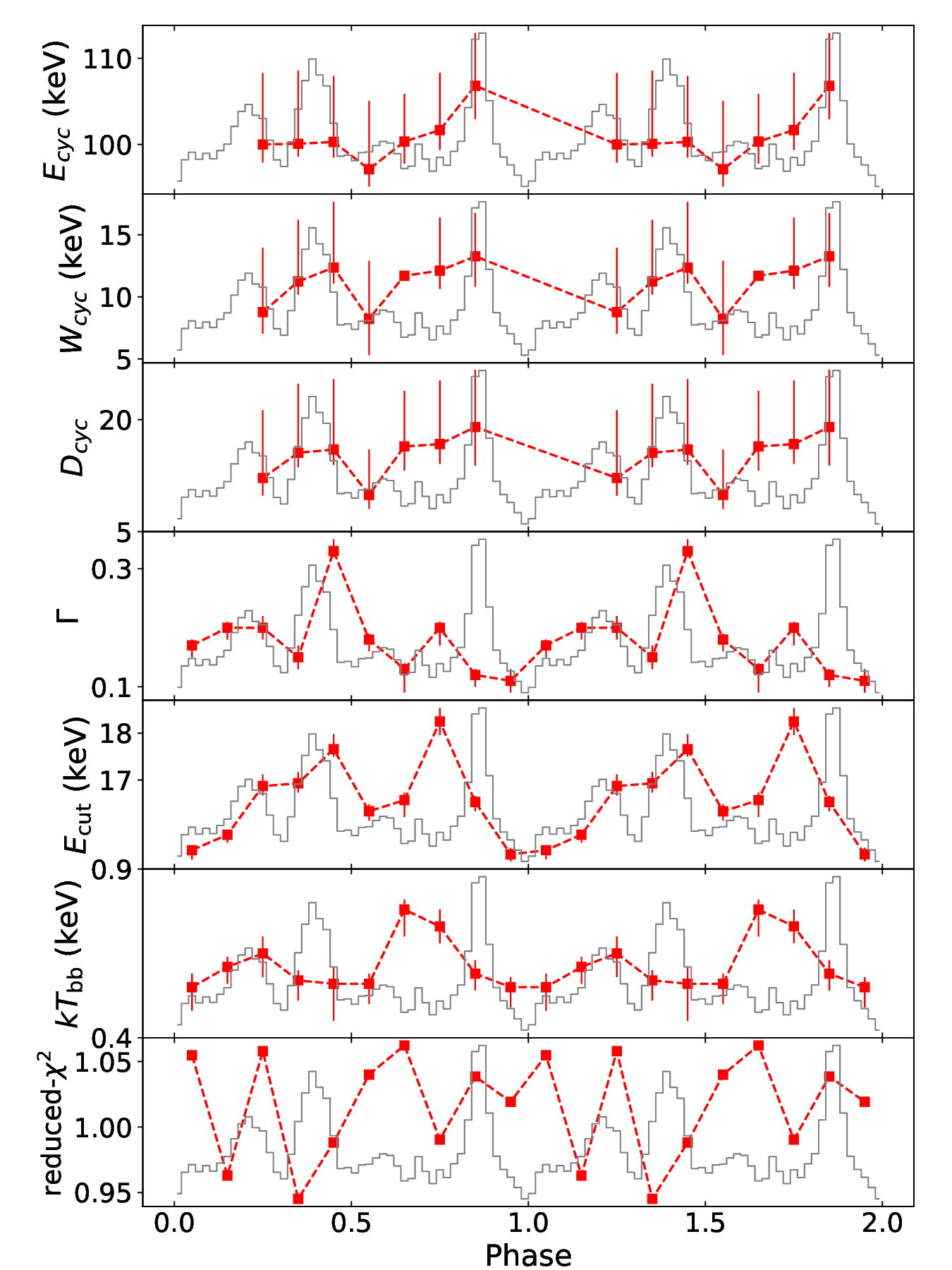}
    \caption{Cyclotron line parameters and spectral parameters shown over the pulse phases. The grey line represents the 30--60 keV pulse profile after binary correction.}
    \label{fig:Par_Phase}
\end{figure}

We also applied other phenomenological models, such as the Fermi–Dirac cut-off power law and the positive and negative power laws multiplied by an exponential cut-off continuum model. These models were also able to fit the spectrum, but slightly worse so than the two models mentioned above. Therefore, we decided to perform the spectral fitting with the cut-off PL and HIGHECUT models, and the results are shown in Table \cref{tab:pars}. Figure \cref{fig:spectrum} (panel (b)) shows the residuals for the two best-fit models of the cut-off PL (left plots) and HIGHECUT model (right plots) separately. We found that the obtained reduced $\chi^2$ for the best fits using the cut-off PL model is given by 1.08 for 1348 d.o.f. As shown in Fig. \cref{fig:spectrum} (b), however, clear residuals between 90 and 110 keV are still visible and may indicate an additional broad CRSF component. Therefore, we added a multiplicative absorption Gaussian line, which is extensively used, for instance, gabs in XSPEC, to model the observed CRSF. The best-fitting results are shown in panel (a) of Fig. \cref{fig:spectrum} and indeed improve the fits (reduced-$\chi^{2}$/d.o.f = 1.04/1345). As listed in Table \cref{tab:pars}, the centroid energy of the CRSFs for the best fit is to be found at $\sim$103 keV. The width and depth of the CRSF are about $11.7_{-2.8}^{+3.7}$ keV and $14.2_{-3.6}^{+3.5}$~keV, respectively. To determine the significance of the CRSF line, we simulated 10000 spectra using the command \textit{simftest} , and we fitted the simulated spectra using the best model without and with the CRSF component. We then calculated the $\Delta \chi^2$ from the best fits. A histogram of the $\Delta \chi^2$ values generated by the simulated data is presented in Fig. \cref{fig:simul}, where the observed $\Delta \chi^2$ from the real data is plotted as the dashed red line. We found that the observed $\Delta \chi^2$ of 52.0 is greater than the $\Delta \chi^2$ values obtained in the simulations, indicating an extremely low F-test probability and corresponding to a significance greater than 3.9 $\sigma$. Because the additional cyclotron line has three free parameters, we expect the histogram of the $\Delta \chi^2$ to follow a $\chi^2$ distribution with three degrees of freedom, which can be used to accurately estimate the statistical significance. As shown in Fig. \cref{fig:simul}, the 3 d.o.f. $\chi^2$ distribution (represented by the orange line) is consistent with the histogram distribution when $\Delta \chi^2$>0. Thus, the probability under the null hypothesis without the absorption component is $\sim$1.5$\times 10^{-11}$, which means that the cyclotron line is detected at a statistical significance of $\sim$7$\sigma$. The background spectra for LE, ME, and HE are shown in Fig. \cref{fig:back}. We found that the background shape dominates above approximately 110 keV and remains very smooth around the CRSF energy. This confirms that our detection of the CRSF is robust and not influenced by background spectra. We still note that because the spectrum becomes background-dominated around 110 keV, which already borders the CRSF (for a 12 keV width as in Table \cref{tab:pars}), the constraints on the CRSF parameters, especially the width and depth, probably depend on the confidence in the background spectrum. We note that the CRSF line is hardly detected or not prominent in individual (phase-averaged) observations. For instance, in observation P0114709012, fitting the spectra with the CRSF might reveal its presence, but it yielded a low significance. This suggests that the statistics of a single observation are generally insufficient to significantly detect the CRSFs in most cases, and more high-quality statistical observations are needed. Additionally, the 100 keV CRSF was detected only near the peak luminosities, and stacking the remaining observations did not yield a significant detection due to the lower luminosities. We also used the cyclabs model in XSPEC to describe the CRSFs (see Table \cref{tab:pars}), and the line energy is slightly lower compared to the gabs model. To avoid the influence on the choice of the continuum models, the right plots of Fig. \cref{fig:spectrum} show fits of the HIGHECUT together with the gabs model, and the best-fit parameters, also including the HIGHECUT with cyclabs models, are listed together in Table \cref{tab:pars}. These parameters, especially the cyclotron line, are consistent with those in the cut-off PL model.

\subsubsection{Phase-resolved spectroscopy} \label{sec:resolved}

In order to study the variations in the CRSFs and other spectral parameters, we carried out phase-resolved spectroscopy of 2S~1417-624 using the same observations as chosen for the averaged spectra. To perform the phase-resolved analysis, we first divided the phase into ten bins to produce the spectra and response matrices using the Insight-HXMT software. The phase-resolved spectra in every phase bin can be fitted well by the cut-off PL with gabs model. We fixed $N_H$ to the value obtained in the phase-averaged fitting (Table \cref{tab:pars}). The best-fitting phase-resolved spectral parameters are shown in Fig. \cref{fig:Par_Phase}, along with the pulse profile at the 30--60 keV. We confirm the CRSFs in the different pulse phases. In the phase of 0.6--0.7, the CRSFs are not constrained, and we fixed the line width to the averaged value of 11.7. For the phases 0.9--1.2 bins, we did not significantly detect the CRSFs due to the poor statistics.

As shown in Fig. \cref{fig:Par_Phase}, the energy of the CRSF line is found to be in the range of 97--107 keV and has a maximum value in the narrow peak phase of 0.8--0.9. The change in the line energy with pulse phase is up to 10 keV. This variation could be explained by a simple model of the accretion column proposed by \cite{2018A&A...620A.153F}, who considered the height of the line-forming region, the velocity of the accreting materials, and the relativistic effects. The width, $W$, of the CRSFs varies from 8 keV to 13 keV, and the depth, $D$, of the line varies between a value of 10--20 throughout the phases. The photon index $\Gamma$ with an average value of 0.2 varies in the range 0.1 to 0.3 over the pulse,  with a minimum value around phase 0.9, similar to the variation of $\Gamma$ obtained from the phase-resolved spectroscopy at the peak 2018 giant outburst reported by \cite{2019MNRAS.490.2458G} with NuSTAR observation. The cut-off energy $E_{\rm cut}$ behaves similarly as $\Gamma$ over the phase. The blackbody temperature $kT_{\rm bb}$ also varies over the phases. The CRSF is clearly detected in some phase bins of the pulse-resolved spectra that correspond to the peak of the pulse profile, indicating an evident phase dependence of the CRSF energy, which could explain why the CRSF line in individual observations may become diluted when it is averaged.

\section{Discussion} \label{sec:discussion}

We have reported a detailed spectral analysis for 2S~1417-624 with the observations made by Insight-HXMT in the 2018 outburst. We obtained a photon index of $\sim$0.21, a cut-off energy of $\sim$17 keV, and a temperature of $\sim$0.6 keV. These continuum parameters are consistent with those reported in previous works. In addition,
we robustly detected a high-energy cyclotron line around 100 keV with a significance of $\sim$7$\sigma$. This line was only prominent around the peak luminosity of the 2018 outburst. The choice of continuum model does not significantly affect the line energy. The width of the cyclotron line is large, consistent with the higher energy of the CRSFs, like in other sources \citep{2002ApJ...580..394C}. In the phase-averaged spectra, we did not see the residuals at 50 keV. To confirm the possible existence of the absorption feature around 50 keV, we added an absorption line with the energy ﬁxed at 50 keV and the width ﬁxed at 6 keV and obtained a line depth of $< \sim$1 keV. Thus, this absorption line at 100 keV is likely the fundamental CRSF. The magnetic field is given by 
\begin{equation}
B \approx E_{\mathrm{cyc}} \times \frac{(1+z)}{11.57 \ \mathrm{keV}},
\end{equation}
where $z$ is the gravitational redshift, and $B$ is in units 10$^{12}$ Gauss. The estimated magnetic field of the neutron star surface is $\sim$9 $\times$ $10^{12}$ G (we ingored the effect of $z$), which is comparable to the value derived by \cite{2020MNRAS.491.1851J}.

The critical luminosity $L_{\text {crit}}$ is crucial for studying the accretion state transition. Previous results \citep{2018MNRAS.479.5612G} found that the pulse fraction of the pulsar was anti-correlated with the source luminosity during the giant outburst in 2009. They also noted the luminosity dependence of the pulse profiles and interpreted it as the transition from the sub-critical to super-critical regime. This transition is connected to the switch of the emission pattern from pencil to fan beams. According to \citep{2012A&A...544A.123B}, we can estimate the critical luminosity using the magnetic field $B$, 
\begin{equation}
\begin{aligned}
L_{\text {crit}}\simeq 1.49 \times 10^{37} \ \mathrm{erg} \ \mathrm{s}^{-1} \times B^{16 / 15},
\end{aligned}
\end{equation}
for the typical parameters of neutron stars: ${M_{\rm NS}}$ =$1.4 M_{\odot}$, ${R_{\rm NS}}$ =10 km. We estimated the expected $L_{\text {crit}}\sim 1.5 \times 10^{38} \ \mathrm{erg} \ \mathrm{s}^{-1}$, which is large compared to the observed luminosity when we adopt the distance of 11 kpc \citep{1984ApJ...276..621G}. The observations performed by Insight-HXMT in the phase-averaged spectroscopy are at an average flux level of 7.89 $\times$ $10^{-9}$ erg cm$^{-2}$ s$^{-1}$ in 2--140 keV range. We can estimate the flux in different observations using the conversion factor of 1.01 $\times \ 10^{-7}$ erg cts$^{-1}$, calculated by the flux divided by the averaged counts rate of $\sim$0.078 in the energy band of 15--50 keV by \textit{Swift}/BAT, which is comparable to the factor determined by \cite{2020MNRAS.491.1851J}. We note that the peak flux at MJD 58240.94, calculated by this factor, is about 8.31 $\times \ 10^{-9}$ erg cm$^{-2}$ s$^{-1}$, consistent with the value of 8.24 $\times \ 10^{-9}$ erg cm$^{-2}$ s$^{-1}$ in the range of 1--79 keV obtained by \cite{2019MNRAS.490.2458G} using NuSTAR at the same time. \cite{2018MNRAS.479.5612G} concluded that the transition of the pulse profiles from double-peaked to triple-peaked during the 2009 outburst may occur in the (3--4) $\times$ $10^{-9}$ erg cm$^{-2}$ s$^{-1}$ flux range, which is consistent with our observations of an intermediate flux level of (4.5--6.5) $\times \ 10^{-9}$ erg cm$^{-2}$ s$^{-1}$ obtained in the 1--10 keV pulse profiles with Insight-HXMT. Our new results for the evolution of the pulse profiles with flux, particularly in high energy bands, reveal that the critical flux is around 7.0 $\times \ 10^{-9}$ erg cm$^{-2}$ s$^{-1}$, in which the pulse profiles changed from a double-peaked shape to a multi-peaked shape. Therefore, assuming a source distance of 11 kpc \citep{1984ApJ...276..621G}, the 2–140 keV critical luminosity can be calculated to be around $10^{38}$ erg s$^{-1}$. However, given the large uncertainty in the distance measurement, the computed critical luminosity may not be accurate. It is possible that we can use the expected critical luminosity and critical flux to constrain the distance. If the expected $L_{\text {crit}}$ is correct, the distance would be located in the range of 13--14 kpc.

On the other hand, the critical luminosity can be predicted by the relation between the CRSF energy and luminosity. In the sub-critical accretion regime, the positive correlation of the CRSF energy luminosity can be seen. In contrast, an anti-correlation between the CRSF energy and luminosity is expected in the supercritical regime. Some sources, such as V0332+53 \citep{2017MNRAS.466.2143D,2018A&A...610A..88V,1990ApJ...365L..59M,2006MNRAS.371...19T} and 1A 0535+262 \citep{2021ApJ...917L..38K}, have shown a clear negative and positive correlation near the critical luminosity. Therefore, more observational studies of the luminosity dependence of the CRSF are required, which can further determine the accretion regime along with the critical luminosity of the source.

\begin{acknowledgements}
We are grateful to the referee for the critical suggestions, which have helped improve the manuscript. This work is supported by the NSFC (12133007) and the National Key Research and Development Program of China (Grant No. 2021YFA0718503). The authors thank the support from the Sino-German (CSC-DAAD) Postdoc Scholarship Program, 2023 (57678375). This work has made use of data from the Insight-HXMT mission, a project funded by China National Space Administration (CNSA) and the Chinese Academy of Sciences (CAS).
\end{acknowledgements}
\bibliographystyle{aa}
\bibliography{refer}

\begin{thebibliography}{40}
\expandafter\ifx\csname natexlab\endcsname\relax\def\natexlab#1{#1}\fi

\bibitem[{{Apparao} {et~al.}(1980){Apparao}, {Naranan}, {Kelley}, \& {Bradt}}]{Apparao1980}
{Apparao}, K.~M.~V., {Naranan}, S., {Kelley}, R.~L., \& {Bradt}, H.~V. 1980, \aap, 89, 249

\bibitem[{{Arnaud}(1996)}]{1996ASPC..101...17A}
{Arnaud}, K.~A. 1996, in Astronomical Society of the Pacific Conference Series, Vol. 101, Astronomical Data Analysis Software and Systems V, ed. G.~H. {Jacoby} \& J.~{Barnes}, 17

\bibitem[{{Bailer-Jones} {et~al.}(2018){Bailer-Jones}, {Rybizki}, {Fouesneau}, {Mantelet}, \& {Andrae}}]{Bailer2018}
{Bailer-Jones}, C.~A.~L., {Rybizki}, J., {Fouesneau}, M., {Mantelet}, G., \& {Andrae}, R. 2018, \aj, 156, 58

\bibitem[{{Basko} \& {Sunyaev}(1976)}]{1976MNRAS.175..395B}
{Basko}, M.~M. \& {Sunyaev}, R.~A. 1976, \mnras, 175, 395

\bibitem[{{Becker} {et~al.}(2012){Becker}, {Klochkov}, {Sch{\"o}nherr}, {Nishimura}, {Ferrigno}, {Caballero}, {Kretschmar}, {Wolff}, {Wilms}, \& {Staubert}}]{2012A&A...544A.123B}
{Becker}, P.~A., {Klochkov}, D., {Sch{\"o}nherr}, G., {et~al.} 2012, \aap, 544, A123

\bibitem[{{Becker} \& {Wolff}(2007)}]{2007ApJ...654..435B}
{Becker}, P.~A. \& {Wolff}, M.~T. 2007, \apj, 654, 435

\bibitem[{{Coburn} {et~al.}(2002){Coburn}, {Heindl}, {Rothschild}, {Gruber}, {Kreykenbohm}, {Wilms}, {Kretschmar}, \& {Staubert}}]{2002ApJ...580..394C}
{Coburn}, W., {Heindl}, W.~A., {Rothschild}, R.~E., {et~al.} 2002, \apj, 580, 394

\bibitem[{{Doroshenko} {et~al.}(2017){Doroshenko}, {Tsygankov}, {Mushtukov}, {Lutovinov}, {Santangelo}, {Suleimanov}, \& {Poutanen}}]{2017MNRAS.466.2143D}
{Doroshenko}, V., {Tsygankov}, S.~S., {Mushtukov}, A.~A., {et~al.} 2017, \mnras, 466, 2143

\bibitem[{{Epili} {et~al.}(2017){Epili}, {Naik}, {Jaisawal}, \& {Gupta}}]{2017MNRAS.472.3455E}
{Epili}, P., {Naik}, S., {Jaisawal}, G.~K., \& {Gupta}, S. 2017, \mnras, 472, 3455

\bibitem[{{Ferrigno} {et~al.}(2016){Ferrigno}, {Pjanka}, {Bozzo}, {Klochkov}, {Ducci}, \& {Zdziarski}}]{2016A&A...593A.105F}
{Ferrigno}, C., {Pjanka}, P., {Bozzo}, E., {et~al.} 2016, \aap, 593, A105

\bibitem[{{Finger} {et~al.}(1996){Finger}, {Wilson}, \& {Chakrabarty}}]{Finger1996}
{Finger}, M.~H., {Wilson}, R.~B., \& {Chakrabarty}, D. 1996, \aaps, 120, 209

\bibitem[{{F{\"u}rst} {et~al.}(2018){F{\"u}rst}, {Falkner}, {Marcu-Cheatham}, {Grefenstette}, {Tomsick}, {Pottschmidt}, {Walton}, {Natalucci}, \& {Kretschmar}}]{2018A&A...620A.153F}
{F{\"u}rst}, F., {Falkner}, S., {Marcu-Cheatham}, D., {et~al.} 2018, \aap, 620, A153

\bibitem[{{Grindlay} {et~al.}(1984){Grindlay}, {Petro}, \& {McClintock}}]{1984ApJ...276..621G}
{Grindlay}, J.~E., {Petro}, L.~D., \& {McClintock}, J.~E. 1984, \apj, 276, 621

\bibitem[{{Guo} {et~al.}(2020){Guo}, {Liao}, {Zhang}, {Zhang}, {Tan}, {Song}, {Lu}, {Cao}, {Chang}, {Chen}, {Du}, {Ge}, {Gu}, {Jiang}, {Jin}, {Li}, {Li}, {Li}, {Liu}, {Liu}, {Lu}, {Luo}, {Meng}, {Sun}, {Yang}, {Yang}, {You}, {Zhang}, {Zhao}, \& {Zhang}}]{2020JHEAp..27...44G}
{Guo}, C.-C., {Liao}, J.-Y., {Zhang}, S., {et~al.} 2020, Journal of High Energy Astrophysics, 27, 44

\bibitem[{{Gupta} {et~al.}(2019){Gupta}, {Naik}, \& {Jaisawal}}]{2019MNRAS.490.2458G}
{Gupta}, S., {Naik}, S., \& {Jaisawal}, G.~K. 2019, \mnras, 490, 2458

\bibitem[{{Gupta} {et~al.}(2018){Gupta}, {Naik}, {Jaisawal}, \& {Epili}}]{2018MNRAS.479.5612G}
{Gupta}, S., {Naik}, S., {Jaisawal}, G.~K., \& {Epili}, P.~R. 2018, \mnras, 479, 5612

\bibitem[{{{\.I}nam} {et~al.}(2004){{\.I}nam}, {Baykal}, {Matthew Scott}, {Finger}, \& {Swank}}]{Inam2004}
{{\.I}nam}, S.~{\c C}., {Baykal}, A., {Matthew Scott}, D., {Finger}, M., \& {Swank}, J. 2004, \mnras, 349, 173

\bibitem[{{Ji} {et~al.}(2020){Ji}, {Doroshenko}, {Santangelo}, {G{\"u}ng{\"o}r}, {Zhang}, {Ducci}, {Zhang}, {Ge}, {Qu}, {Chen}, {Bu}, {Cai}, {Cao}, {Chang}, {Chen}, {Chen}, {Chen}, {Chen}, {Chen}, {Cui}, {Cui}, {Deng}, {Dong}, {Du}, {Fu}, {Gao}, {Gao}, {Gao}, {Gu}, {Guan}, {Guo}, {Han}, {Huang}, {Huo}, {Jia}, {Jiang}, {Jiang}, {Jin}, {Kong}, {Li}, {Li}, {Li}, {Li}, {Li}, {Li}, {Li}, {Li}, {Li}, {Li}, {Li}, {Liang}, {Liao}, {Liu}, {Liu}, {Liu}, {Liu}, {Liu}, {Lu}, {Lu}, {Lu}, {Luo}, {Luo}, {Ma}, {Meng}, {Nang}, {Nie}, {Ou}, {Sai}, {Song}, {Song}, {Sun}, {Tan}, {Tao}, {Tuo}, {Wang}, {Wang}, {Wang}, {Wang}, {Wang}, {Wen}, {Wu}, {Wu}, {Wu}, {Xiao}, {Xiao}, {Xiong}, {Xu}, {Yang}, {Yang}, {Yang}, {Yang}, {Yi}, {Yin}, {You}, {Zhang}, {Zhang}, {Zhang}, {Zhang}, {Zhang}, {Zhang}, {Zhang}, {Zhang}, {Zhang}, {Zhang}, {Zhang}, {Zhang}, {Zhang}, {Zhang}, {Zhang}, {Zhao}, {Zhao}, {Zheng}, {Zhou}, {Zhou}, {Zhu}, \& {Zhu}}]{2020MNRAS.491.1851J}
{Ji}, L., {Doroshenko}, V., {Santangelo}, A., {et~al.} 2020, \mnras, 491, 1851

\bibitem[{{Kelley} {et~al.}(1981){Kelley}, {Apparao}, {Doxsey}, {Jernigan}, {Naranan}, \& {Rappaport}}]{1981ApJ...243..251K}
{Kelley}, R.~L., {Apparao}, K.~M.~V., {Doxsey}, R.~E., {et~al.} 1981, \apj, 243, 251

\bibitem[{{Kong} {et~al.}(2021){Kong}, {Zhang}, {Ji}, {Reig}, {Doroshenko}, {Santangelo}, {Staubert}, {Zhang}, {Soria}, {Chang}, {Chen}, {Wang}, {Tao}, \& {Qu}}]{2021ApJ...917L..38K}
{Kong}, L.~D., {Zhang}, S., {Ji}, L., {et~al.} 2021, \apjl, 917, L38

\bibitem[{{Liao} {et~al.}(2020{\natexlab{a}}){Liao}, {Zhang}, {Chen}, {Zhang}, {Jin}, {Chang}, {Chen}, {Ge}, {Guo}, {Li}, {Li}, {Lu}, {Lu}, {Nie}, {Song}, {Yang}, {You}, {Zhao}, \& {Zhang}}]{2020JHEAp..27...24L}
{Liao}, J.-Y., {Zhang}, S., {Chen}, Y., {et~al.} 2020{\natexlab{a}}, Journal of High Energy Astrophysics, 27, 24

\bibitem[{{Liao} {et~al.}(2020{\natexlab{b}}){Liao}, {Zhang}, {Lu}, {Zhang}, {Li}, {Chang}, {Chen}, {Ge}, {Guo}, {Huang}, {Jin}, {Li}, {Li}, {Li}, {Liu}, {Lu}, {Nie}, {Song}, {Wang}, {You}, {Zhang}, {Zhao}, \& {Zhang}}]{2020JHEAp..27...14L}
{Liao}, J.-Y., {Zhang}, S., {Lu}, X.-F., {et~al.} 2020{\natexlab{b}}, Journal of High Energy Astrophysics, 27, 14

\bibitem[{{Lutovinov} \& {Tsygankov}(2009)}]{2009AstL...35..433L}
{Lutovinov}, A.~A. \& {Tsygankov}, S.~S. 2009, Astronomy Letters, 35, 433

\bibitem[{{Makishima} {et~al.}(1990){Makishima}, {Mihara}, {Ishida}, {Ohashi}, {Sakao}, {Tashiro}, {Tsuru}, {Kii}, {Makino}, {Murakami}, {Nagase}, {Tanaka}, {Kunieda}, {Tawara}, {Kitamoto}, {Miyamoto}, {Yoshida}, \& {Turner}}]{1990ApJ...365L..59M}
{Makishima}, K., {Mihara}, T., {Ishida}, M., {et~al.} 1990, \apjl, 365, L59

\bibitem[{{Mandal} \& {Pal}(2022)}]{2022Ap&SS.367..112M}
{Mandal}, M. \& {Pal}, S. 2022, \apss, 367, 112

\bibitem[{{Maraschi} {et~al.}(1976){Maraschi}, {Treves}, \& {van den Heuvel}}]{1976Natur.259..292M}
{Maraschi}, L., {Treves}, A., \& {van den Heuvel}, E.~P.~J. 1976, \nat, 259, 292

\bibitem[{{Mushtukov} \& {Tsygankov}(2022)}]{2022arXiv220414185M}
{Mushtukov}, A. \& {Tsygankov}, S. 2022, arXiv e-prints, arXiv:2204.14185

\bibitem[{{Mushtukov} {et~al.}(2015{\natexlab{a}}){Mushtukov}, {Suleimanov}, {Tsygankov}, \& {Poutanen}}]{2015MNRAS.454.2539M}
{Mushtukov}, A.~A., {Suleimanov}, V.~F., {Tsygankov}, S.~S., \& {Poutanen}, J. 2015{\natexlab{a}}, \mnras, 454, 2539

\bibitem[{{Mushtukov} {et~al.}(2015{\natexlab{b}}){Mushtukov}, {Suleimanov}, {Tsygankov}, \& {Poutanen}}]{2015MNRAS.447.1847M}
{Mushtukov}, A.~A., {Suleimanov}, V.~F., {Tsygankov}, S.~S., \& {Poutanen}, J. 2015{\natexlab{b}}, \mnras, 447, 1847

\bibitem[{{Postnov} {et~al.}(2015){Postnov}, {Gornostaev}, {Klochkov}, {Laplace}, {Lukin}, \& {Shakura}}]{2015MNRAS.452.1601P}
{Postnov}, K.~A., {Gornostaev}, M.~I., {Klochkov}, D., {et~al.} 2015, \mnras, 452, 1601

\bibitem[{{Poutanen} {et~al.}(2013){Poutanen}, {Mushtukov}, {Suleimanov}, {Tsygankov}, {Nagirner}, {Doroshenko}, \& {Lutovinov}}]{2013ApJ...777..115P}
{Poutanen}, J., {Mushtukov}, A.~A., {Suleimanov}, V.~F., {et~al.} 2013, \apj, 777, 115

\bibitem[{{Raichur} \& {Paul}(2010)}]{2010MNRAS.406.2663R}
{Raichur}, H. \& {Paul}, B. 2010, \mnras, 406, 2663

\bibitem[{{Reig}(2011)}]{2011Ap&SS.332....1R}
{Reig}, P. 2011, \apss, 332, 1

\bibitem[{{Serim} {et~al.}(2022){Serim}, {{\"O}z{\"u}do{\u{g}}ru}, {D{\"o}nmez}, {{\c{S}}ahiner}, {Serim}, {Baykal}, \& {{\.I}nam}}]{2022MNRAS.510.1438S}
{Serim}, M.~M., {{\"O}z{\"u}do{\u{g}}ru}, {\"O}.~C., {D{\"o}nmez}, {\c{C}}.~K., {et~al.} 2022, \mnras, 510, 1438

\bibitem[{{Thalhammer} {et~al.}(2021){Thalhammer}, {Bissinger}, {Ballhausen}, {Pottschmidt}, {Wolff}, {Stierhof}, {Sokolova-Lapa}, {F{\"u}rst}, {Malacaria}, {Gottlieb}, {Marcu-Cheatham}, {Becker}, \& {Wilms}}]{2021A&A...656A.105T}
{Thalhammer}, P., {Bissinger}, M., {Ballhausen}, R., {et~al.} 2021, \aap, 656, A105

\bibitem[{{Tsygankov} {et~al.}(2006){Tsygankov}, {Lutovinov}, {Churazov}, \& {Sunyaev}}]{2006MNRAS.371...19T}
{Tsygankov}, S.~S., {Lutovinov}, A.~A., {Churazov}, E.~M., \& {Sunyaev}, R.~A. 2006, \mnras, 371, 19

\bibitem[{{Tsygankov} {et~al.}(2017){Tsygankov}, {Wijnands}, {Lutovinov}, {Degenaar}, \& {Poutanen}}]{2017MNRAS.470..126T}
{Tsygankov}, S.~S., {Wijnands}, R., {Lutovinov}, A.~A., {Degenaar}, N., \& {Poutanen}, J. 2017, \mnras, 470, 126

\bibitem[{{Vybornov} {et~al.}(2018){Vybornov}, {Doroshenko}, {Staubert}, \& {Santangelo}}]{2018A&A...610A..88V}
{Vybornov}, V., {Doroshenko}, V., {Staubert}, R., \& {Santangelo}, A. 2018, \aap, 610, A88

\bibitem[{{Wilms} {et~al.}(2000){Wilms}, {Allen}, \& {McCray}}]{2000ApJ...542..914W}
{Wilms}, J., {Allen}, A., \& {McCray}, R. 2000, \apj, 542, 914

\bibitem[{{Zhang} {et~al.}(2020){Zhang}, {Li}, {Lu}, {Song}, {Xu}, {Liu}, {Chen}, {Cao}, {Bu}, {Chang}, {Chen}, {Chen}, {Chen}, {Chen}, {Chen}, {Cui}, {Cui}, {Deng}, {Dong}, {Du}, {Fu}, {Gao}, {Gao}, {Gao}, {Ge}, {Gu}, {Guan}, {Gungor}, {Guo}, {Han}, {Hu}, {Huang}, {Huo}, {Jia}, {Jiang}, {Jiang}, {Jin}, {Jin}, {Li}, {Li}, {Li}, {Li}, {Li}, {Li}, {Li}, {Li}, {Li}, {Li}, {Li}, {Liang}, {Liao}, {Liu}, {Liu}, {Liu}, {Liu}, {Liu}, {Liu}, {Lu}, {Lu}, {Luo}, {Ma}, {Meng}, {Nang}, {Nie}, {Ou}, {Qu}, {Sai}, {Shang}, {Shen}, {Sun}, {Tan}, {Tao}, {Tuo}, {Wang}, {Wang}, {Wang}, {Wang}, {Wang}, {Wang}, {Wang}, {Wen}, {Wu}, {Wu}, {Wu}, {Xiao}, {Xiong}, {Yan}, {Yang}, {Yang}, {Yang}, {Yi}, {Yuan}, {Zhang}, {Zhang}, {Zhang}, {Zhang}, {Zhang}, {Zhang}, {Zhang}, {Zhang}, {Zhang}, {Zhang}, {Zhang}, {Zhang}, {Zhang}, {Zhang}, {Zhang}, {Zhang}, {Zhang}, {Zhang}, {Zhang}, {Zhang}, {Zhao}, {Zhao}, {Zheng}, {Zhou}, {Zhu}, {Zhu}, {Zhuang}, \& {Insight-HXMT Team}}]{2020SCPMA..6349502Z}
{Zhang}, S.-N., {Li}, T., {Lu}, F., {et~al.} 2020, Science China Physics, Mechanics, and Astronomy, 63, 249502

\end{thebibliography}

\appendix
\onecolumn
\section{Additional material}

\begin{table*}[h]
    \centering
    \footnotesize
    \caption{Observation IDs of Insight-HXMT in the 2018 outburst.}
    \label{tab:ObsIDs}
    \renewcommand\arraystretch{1.5}
    \setlength{\tabcolsep}{1.2mm}{
    \begin{tabular}{l|cccccccc}
    \hline \hline 
    \multirow{2}{*}{ObsID} & \multirow{2}{*}{Time Start (UTC)}  & \multirow{2}{*}{\thead{Exposure time (s) \\ HE/ME/LE}} & \multirow{2}{*}{MJD} & \multirow{2}{*}{\thead{Flux$^a$ \\ (10$^{-9}$ erg cm$^{-2}$ s$^{-1}$)}}  & \multirow{2}{*}{\thead{PF$^b$ \\ (30-60 keV)}} & \multirow{2}{*}{\thead{PF \\ (10-30 keV)}} & \multirow{2}{*}{\thead{PF \\ (1-10 keV)}}\\ \\
    \hline 
P0114709001 & 2018-04-13 &      2055/1710/1256  & 58221.62 &    $4.50 \pm 0.20$ &         $0.45 \pm 0.02$ &       $0.32 \pm 0.03$ &       $0.30 \pm 0.05$ \\ 
P0114709002 & 2018-04-14 &      340/1020/299    & 58222.41 &    $4.47 \pm 0.28$ & $0.43 \pm 0.05$ &       $0.35 \pm 0.04$ &       $0.33 \pm 0.09$ \\
P0114709003 & 2018-04-15 &      4492/4470/3411  & 58223.54 &    $4.71 \pm 0.38$ & $0.39 \pm 0.01$ &       $0.24 \pm 0.02$ &       $0.23 \pm 0.03$ \\
P0114709004 & 2018-04-16 &      3648/3239/2098  & 58224.15 &    $5.20 \pm 0.36$ & $0.37 \pm 0.01$ &       $0.28 \pm 0.02$ &       $0.28 \pm 0.03$ \\
P0114709005 & 2018-04-17 &      2778/2670/1436  & 58225.14 &    $5.93 \pm 0.31$ & $0.37 \pm 0.01$ &       $0.25 \pm 0.02$ &       $0.20 \pm 0.04$ \\
P0114709006 & 2018-04-18 &      2771/2610/1496  & 58226.13 &    $6.33 \pm 0.29$ & $0.36 \pm 0.01$ &       $0.28 \pm 0.02$ &       $0.27 \pm 0.04$ \\
P0114709007 & 2018-04-19 &      4693/4050/3171  & 58227.52 &    $6.60 \pm 0.28$ & $0.31 \pm 0.01$ &       $0.27 \pm 0.01$ &       $0.24 \pm 0.02$ \\
P0114709008 & 2018-04-20 &      629/2190/1316   & 58228.31 &    $6.89 \pm 0.28$ & $0.27 \pm 0.03$ &       $0.28 \pm 0.02$ &       $0.29 \pm 0.04$ \\
P0114709009 & 2018-04-21 &      2186/1980/952   & 58229.12 &    $7.00 \pm 0.28$ & $0.31 \pm 0.01$ &       $0.28 \pm 0.02$ &       $0.27 \pm 0.04$ \\
P0114709010 & 2018-04-21 &      3247/2790/1518  & 58230.04 &    $7.61 \pm 0.31$ & $0.28 \pm 0.01$ &       $0.28 \pm 0.02$ &       $0.24 \pm 0.03$ \\
P0114709011 & 2018-04-23 &      1223/2520/837   & 58231.31 &    $7.20 \pm 0.31$ & $0.35 \pm 0.02$ &       $0.26 \pm 0.02$ &       $0.24 \pm 0.04$ \\
P0114709012 & 2018-04-24 &      3103/2940/1316  & 58232.37 &    $7.65 \pm 0.30$ & $0.26 \pm 0.01$ &       $0.30 \pm 0.02$ &       $0.29 \pm 0.03$ \\
P0114709013 & 2018-04-25 &      2466/3360/2304  & 58233.48 &    $7.78 \pm 0.30$ & $0.29 \pm 0.01$ &       $0.25 \pm 0.01$ &       $0.26 \pm 0.02$ \\
P0114709014 & 2018-04-27 &      3687/3419/1473  & 58235.88 &    $7.62 \pm 0.31$ & $0.29 \pm 0.01$ &       $0.26 \pm 0.01$ &       $0.24 \pm 0.03$ \\
P0114709015 & 2018-04-28 &      945/1800/172    & 58237.09 &    $7.70 \pm 0.31$ & $0.28 \pm 0.02$ &       $0.27 \pm 0.02$ &       $0.33 \pm 0.08$ \\
P0114709016 & 2018-04-30 &      475/2700/1697   & 58238.48 &    $7.85 \pm 0.36$ & $0.34 \pm 0.03$ &       $0.28 \pm 0.02$ &       $0.26 \pm 0.03$ \\
P0114709017 & 2018-05-01 &      4545/4500/2094  & 58239.68 &    $7.67 \pm 0.32$ & $0.27 \pm 0.01$ &       $0.29 \pm 0.01$ &       $0.24 \pm 0.02$ \\
P0114709018 & 2018-05-02 &      2361/2430/636   & 58240.94 &    $8.31 \pm 0.32$ & $0.27 \pm 0.01$ &       $0.28 \pm 0.02$ &       $0.29 \pm 0.04$ \\
P0114709019 & 2018-05-03 &      3184/3030/1182  & 58241.85 &    $8.17 \pm 0.31$ & $0.27 \pm 0.01$ &       $0.29 \pm 0.01$ &       $0.28 \pm 0.03$ \\
P0114709020 & 2018-05-04 &      4719/4590/2054  & 58242.66 &    $7.95 \pm 0.30$ & $0.28 \pm 0.01$ &       $0.26 \pm 0.01$ &       $0.25 \pm 0.02$ \\
P0114709021 & 2018-05-05 &      3423/3329/924   & 58243.77 &    $7.95 \pm 0.31$ & $0.28 \pm 0.01$ &       $0.26 \pm 0.01$ &       $0.28 \pm 0.04$ \\
P0114709022 & 2018-07-04 &      4359/6080/897   & 58303.37 &    $2.16 \pm 0.25$ & $0.55 \pm 0.03$ &       $0.40 \pm 0.02$ &       $0.29 \pm 0.06$ \\
P0114709023 & 2018-07-05 &      1079/3000/478   & 58304.26 &    $2.18 \pm 0.29$ & $0.57 \pm 0.06$ &       $0.43 \pm 0.03$ &       $0.51 \pm 0.12$ \\
P0114709025 & 2018-07-06 &      4173/4110/1436  & 58305.57 &    $1.99 \pm 0.26$ & $0.56 \pm 0.03$ &       $0.43 \pm 0.03$ &       $0.34 \pm 0.06$ \\
P0114709026 & 2018-07-07 &      2225/2190/897   & 58306.76 &    $1.80 \pm 0.24$ & $0.73 \pm 0.06$ &       $0.41 \pm 0.04$ &       $0.38 \pm 0.08$ \\
P0114709027 & 2018-07-08 &      970/2130/837    & 58307.82 &    $1.80 \pm 0.18$ & $0.69 \pm 0.08$ &       $0.52 \pm 0.05$ &       $0.49 \pm 0.08$ \\
P0114709028 & 2018-07-09 &      1278/2909/778   & 58308.89 &    $1.97 \pm 0.19$ & $0.45 \pm 0.05$ &       $0.38 \pm 0.04$ &       $0.36 \pm 0.08$ \\
P0114709029 & 2018-07-10 &      1098/1710/658   & 58309.74 &    $1.99 \pm 0.18$ & $0.65 \pm 0.07$ &       $0.42 \pm 0.05$ &       $0.47 \pm 0.09$ \\
P0114709030 & 2018-07-11 &      1916/1800/778   & 58310.66 &    $1.44 \pm 1.06$ & $0.86 \pm 0.08$ &       $0.46 \pm 0.05$ &       $0.36 \pm 0.07$ \\
    \hline
\end{tabular}
}
\begin{tablenotes}
\item $^a$ The unabsorbed flux of 2S~1417-624 is computed from the count rate history provided by \textit{SWIFT}/BAT by multiplying a conversion factor of 1.01 $\times \ 10^{-7}$ erg cts$^{-1}$ in 2–140~keV range (see Sect. \cref{sec:discussion}). 
\item $^b$ The pulse fraction (PF=($I_{max}$-$I_{min}$)/($I_{max}$+$I_{min}$)) is calculated from the background-subtracted
light curves.
\end{tablenotes}
\end{table*}

\begin{figure*}
    \centering
    \includegraphics[width=.99\textwidth]{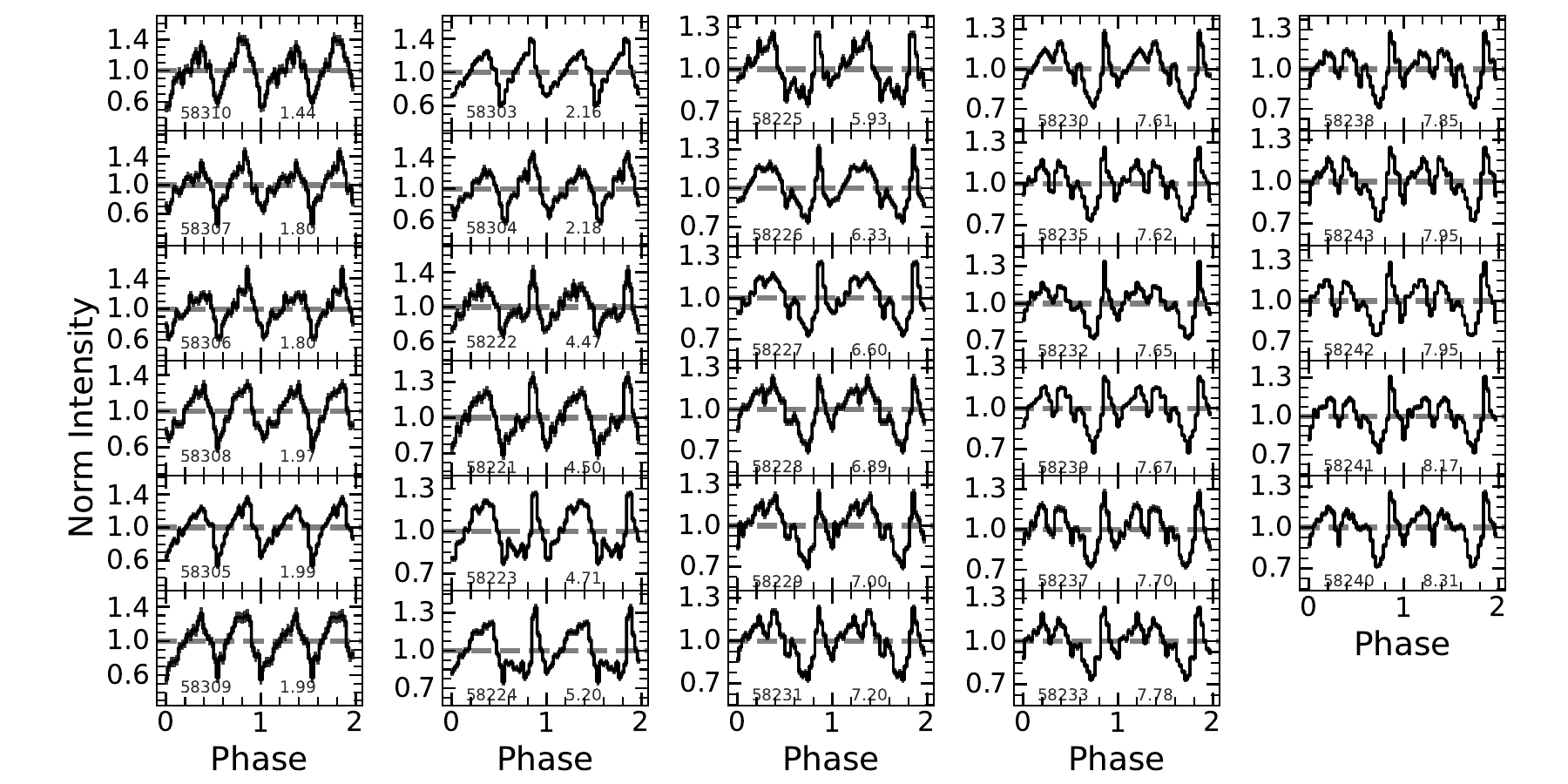}
    \caption{Pulse proﬁles in the energy range of 10–30 keV from Insight-HXMT/ME data at different time (in MJD, left hand of each panel) and flux (in $10^{-9}$ erg cm$^{-2}$ s$^{-1}$, right hand of each panel) during the 2018 giant outburst.}
    \label{fig:pulse_ME}
\end{figure*}

\begin{figure*}
    \centering
    \includegraphics[width=.99\textwidth]{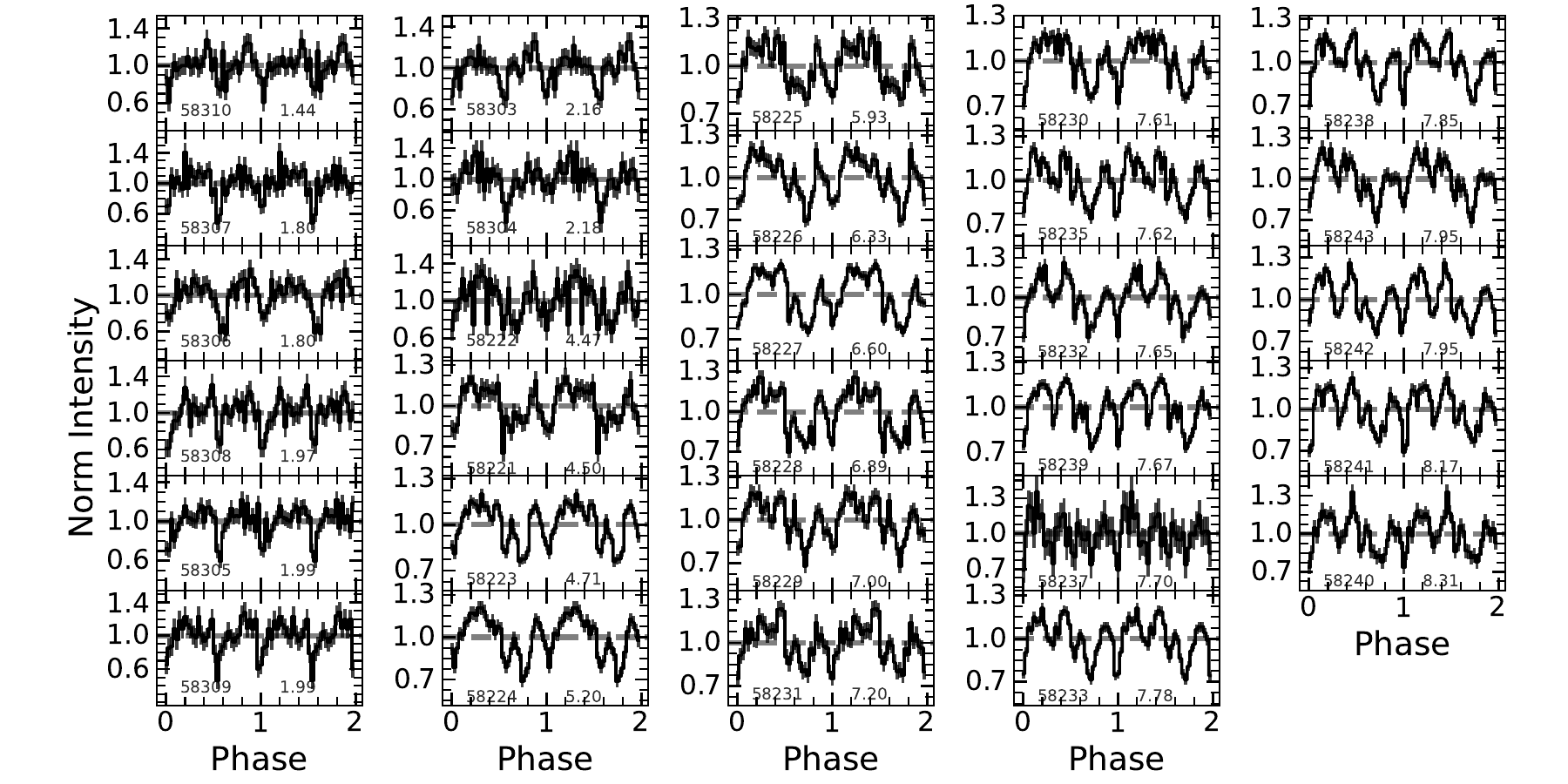}
    \caption{Pulse proﬁles in the energy range of 1–10 keV from Insight-HXMT/LE data at different time (in MJD, left hand of each panel) and flux (in $10^{-9}$ erg cm$^{-2}$ s$^{-1}$, right hand of each panel) during the 2018 giant outburst.}
    \label{fig:pulse_LE}
\end{figure*}
\end{document}